\documentstyle [12pt]{article}
\begin{document}
\newcommand{\vm}{\vspace{0.2cm}}
\newcommand{\vl}{\vspace{0.4cm}}

\title{p-Adic  description of Higgs mechanism III: calculation of elementary
particle  masses }
\author{Matti Pitk\"anen\\
Torkkelinkatu 21 B 39,00530, Helsinki, FINLAND}
\date{6.10. 1994}
\maketitle

\newpage
\begin{center}
Abstract
\end{center}

\vm

 This paper  belongs to  the series devoted to the calculation of particle
masses in the framework of p-adic conformal field theory limit of
Topological GeometroDynamics. In  paper II the general formulation of
p-adic Higgs mechanism was given. In this paper    the calculation of the
fermionic and bosonic masses is carried out. The calculation of the
masses necessitates the evaluation of degeneracies for states  as a
function of  conformal weight in certain tensor product of
 Super Virasoro algebras. The masses are very sensitive to the degeneracy
ratios: Planck mass
 results unless the ratio for the degeneracies for first excited states
and massless states is  an integer multiple of 2/3.  For leptons, quarks
and gauge bosons this miracle
 occurs. The main deviation from standard model is the prediction of
light  color excited  leptons and quarks as well as colored exotic bosons.
 Higgsis absent from the spectrum as is also graviton: the latter is in
accordance with the basic assumptions of p-adic field theory limit of TGD.

\newpage
\tableofcontents

\newpage

\section{Introduction}

This is the  third paper in the series devoted to the p-adic description of
 Higgs mechanism in TGD \cite{TGD,padTGD} (for p-adic numbers see for
instance
 \cite{padrev}).
  Concerning the general background reader is suggested to read the
introduction
 of the first
  paper, where general formulation of p-adic conformal field theory limit
was
 proposed and  predictions are summarized. The general theory of Higgs
mechanism was  described
 in previous paper.

\vm

  The calculation of masses of leptons, quarks  and gauge bosons is
carried out  in this paper applying p-adic thermodynamics.   The
calculation of the  masses
 boils down to the evaluation of degeneracies for states  as a function of
conformal weight in
 certain tensor product of Super Virasoro algebras.  There are  some
delicacies associated with the  norm of the states since inner product is
defined as the massless limit of the inner product  associated with the
symmetry broken Super Virasoro representation. The thermal expectation
values
 for fermionic masses are of order Planck mass for physical values of
p-adic prime $p$ unless the  ratio for  the degeneracies of $M^2=3/2$
Planck mass states  and $M^2=0$ states is integer
 multiple of $2/3$. It turns out  that  this miracle occurs for leptons
and quarks   as well as  certain colored excitations of leptons and U
type  quarks. The physical consequences of the  exotic light leptons and
quarks are considered in the fifth paper of the series.    Although there
are hundreds of exotic bosons there are  no massless noncolored exotic
 bosons so that no new long range forces are predicted. Higgs particle is
absent from
 the spectrum  as is also graviton: the latter in accordance with the basic
assumptions about p-adic conformal
 field theory limit.

\vm

The predictions for lepton and gauge bosons masses agree surprisingly well
with known experimental masses.  Errors are below one per cent except for
$Z^0$ boson for which mass is $10$ per cent too large. The reason is too
large value $sin^2(\theta_W)=3/8$ for Weinberg angle:  in the third paper
it is shown that
 inclusion of Coulombic corrections and topological mixing effects of
leptons leads to a
 correct prediction for gauge boson masses.    One can safely conclude
that the results of the
 calculation verify the
 essential correctness of  both TGD and its p-adic conformal field theory
limit at quantitative level.    The detailed analysis and application  of
the results to derive  information on of hadron masses is left to the
fourth and fifth papers of the series.

 \vm

 Since the masses are very sensitive to the degeneracy ratios and  since
modulo  arithmetics as well as the canonical correspondence between p-adic
and real numbers is essentially
 involved,  the details of the calculations are given   in the form of
appendices are listed so  that interested reader can check  them.

\section{Calculation of elementary  fermion and boson masses}

In the sequel the  calculations of elementary fermion  and  elementary
boson masses are described at general level. The details of the
calculations are left to appendix.

\subsection{Modular contribution to the mass of  elementary particle}

 The thermal independence of cm and modular degrees of freedom implies
 that mass squared for elementary particle   is sum of cm and modular
contributions:

\begin{eqnarray}
M^2&=& M^2(cm)+M^2(mod)
\end{eqnarray}

\noindent   The general form of the  modular contribution   should
 be derivable from p-adic partition function for conformally invariant
degrees of freedom associated with boundary components. The general form
of vacuum state functionals as modular invariant functions of Teichmuller
parameters   was derived in \cite{TGD} and the square of the elementary
particle vacuum functional can be identified as partition function.  Even
theta functions serve as basic building blocks and the functionals are
proportional to the product of all even theta functions  and their complex
conjugates.   The number of theta functions for genus $g>0$ is given by

\begin{eqnarray}
N(g) &=& 2^{g-1}(2^g+1)
\end{eqnarray}

\noindent  One has $N(1)=3$  for muon and  $N(2)=10$ for $\tau$. \\ e)
Single theta function is analogous to partition function. This implies
that the modular contribution  to mass squared must be proportional to
$2N(g)$.  The factor two follows from the presence of both  theta functions
and their conjugates in partition function.  \\ f)  The factorization
properties of the vacuum functionals imply that  handles behave
effectively as particles. For example,  at the limit,  when the surface
splits into two pieces with $g_1$ and $g-g_1$ handles the partition
function reduces to product of $g_1$ and $g-g_1$ partition functions. This
implies that the contribution to mass squared is proportional to the genus
of the surface.  Altogether one has

\begin{eqnarray}
M^2(mod,g)  &=& k(F) k(mod) 2N(g)g \frac{M^2_0}{p}\nonumber\\
k(F)&=&3/2\nonumber\\
k(mod)&=&1
\end{eqnarray}

\noindent Here $k(mod)$ is some integer valued constant (in order to avoid
 Planck mass) to be determined.   $k(mod)=1$ turns out to be the correct
choice
 for this parameter.   The value of the constant  $k(F)$  in $M^2=
k(F)L_0+...$  is
 analogous to the string tension parameter and  is determined uniquely
from the requirement that actual tachyons are absent. Also the
requirements that photons are massless, that  charged leptons are light
and that charged lepton mass ratios are predicted correctly,  fix the
value of this parameter to $k(B)=k(F)=3/2$.

\vm

Summarizing,  modular contribution to the mass of
 elementary particle (also boson!)  belonging to $g+1$:th generation
reads as

\begin{eqnarray}
M^2(mod)&=& 0 \ for \ e, \nu_e, u,d \nonumber\\
M^2(mod)&=& 9 \frac{M^2_0}{p(X))}\ for \ X=\mu, \nu_{\tau},c,s\nonumber\\
M^2(mod)&=& 60 \frac{M^2_0}{p(X)}\ for \   X=\tau, \nu_{\tau},t,b
\end{eqnarray}

\noindent  The formula can be tested also in hadronic context.

\vm

The higher order modular  contributions to mass squared   are completely
negligible  if the degeneracy of massless state is $D(0,mod,g)=1$ in
modular degrees of freedom as is in fact required by $k(mod)=1$. The
absence of vacuum degeneracy is natural assumption and  guarantees
lightness of boundary component for $g<3$.  This implies that second order
contribution to mass squared is uniquely determined by cm contribution.

\subsection{ Calculation of cm  contribution to fermion mass}

Fermion masses are assumed be sums of cm and modular contributions
$M^2=M^2(cm) +M^2(mod)$.
 'cm' refers to the cm of boundary component.   The general form of modular
'contribution is determined completely by general arguments. The
calculation of the  cm contribution thermodynamically involves
thermodynamics for Ramond type generalized spinors and in principle
calculation is straightforward p-adic  generalization of ordinary
thermodynamics.

 \vm

The possibility to describe  Higgs mechanism thermally sets strong
constraints on the theory.  \\
 a) $M^4$ and $CP_2$ degrees of freedom can be described using generalized
$H$-spinor, which can be constructed as four-fold tensor products of
$so(4)$ Super Virasoro representations.  In $su(3)$' degrees of freedom
$N-S$ type representation is assumed for leptons.  This is possible since
one can generalize the expression for  super supper symmetry generators of
Ramond type appropriately.  The Ramond representation in $U(1)$ degrees of
freedom implies doubling of degeneracies. \\
 b) It turns out that  $h=-5/2$ is the only possible value vacuum weight in
case  of neutrinos and $h=-3/2$ in case of charged leptons. The presence of
tachyon is analogous to the presence of tachyonic Higgs particle (around
 symmetry nonbroken vacuum) in gauge theories. Tachyonic ground state
implies  vacuum degeneracy analogous to the degeneracy of symmetry broken
vacua in gauge theories.\\ c) Tachyons can be eliminated either by direct
constraint on  spectrum or by defining the negative powers of $p$ coming
from tachyonic states as the limit  $p^{-n}\equiv lim_{N\rightarrow \infty
} p^{n(p-1)(1+p+p^2+...+p^N)} =0$.  Their elimination is not absolutely
necessary since tachyonic states satisfying G-parity rule are actually
particles with real mass for physically interesting primes $p$ due to the
special properties of p-adic square root (square roots of $-3/2$ and $-3$
are p-adically real).  Furthermore, even states for which energy $M$ is
imaginary for the particle at rest allow states for which the components
of four momentum are p-adically real.   \\ d) Electroweak mass splitting is
described by assuming that vacuum weight depends on isospin  of the
fermion  via the formula

\begin{eqnarray}
h(\nu)=h(U)&=& -\frac{5}{2}\nonumber\\
h(L^-)=h(D)&=&-\frac{3}{2}
\end{eqnarray}

\noindent  Since K\"ahler charge $ Q_K=\pm 1$ increases vacuum
 weight by $Q_K^2/2=1/2$ one must use operators of weight
 $\Delta= -3/2-I_3+n$ to create  $M^2=n$ states and thermal expectations
for charged leptons and neutrinos are  different. \\ e)  Only integer
excitations of $L_0$ are  allowed.    For fermions half integer excitation
transforms quark  into lepton and vice versa and is excluded by triality
rule of Quantum TGD (quarks and leptons correspond to triality one and
zero representations of color group).  The rule is just the G-parity rule
of string models and in present case is necessary to exclude actual
tachyons from spectrum. \\
  f)    No thermal mixing of different vectorial isospins  takes place and
thermalization occurs in Super Virasoro degrees of freedom only.
Temperature
 parameter ($1/T$ is integer) is assumed to be $T=1$ for fermions.   The
assumption $k(F)=3/2$ is necessary for the elimination of tachyons by
making their masses actually p-adically real and  for prediction of
correct leptonic mass ratios.   \\ g)  The thermal expectation for  p-adic
mass squared can be expressed in  terms of degeneracies  $D(i)$  of
$M^2=0,3/2,3$ states in extremely good approximation as

\begin{eqnarray}
M^2(cm)&=& k(F)\frac{D(3/2)}{D(0)}p +k(F)\frac{(2D(3)-\frac{D(3/2)^2}{D(0)})
}{D(0)}p^2 \nonumber\\ \
\end{eqnarray}

\noindent and the real counterpart of the mass squared is easily obtained
from this expression using canonical correspondences between p-adic and
real numbers.  The task is to calculate the degeneracies. \\ h) If second
order contribution to mass is  written in the form  $(X/64)p^2$  to mass
squared correspond to small integer $X$  then  that ground state
degeneracy must be $D=64$ for fermions  and $D=16$ for bosons or more
generally a power of $2$ not larger than $64$.   It turns out that this
condition is not  realized. \\ i) The condensation levels for charged
lepton families are assumed to be
 $k=127, 113$ and $k=107$ in accordance with the hypothesis about the
importance of primes near prime powers of $2$.  A somewhat puzzling result
is that  the value of $k$ for boundary component is same as for the
interior:  this is
 true   not only for leptons but also for $u,d,s$ quarks and  hadrons
($k=107$ for both).  Since boundary contribution and interior
contributions in principle correspond to two different condensation
levels  this makes sense only provided  primary and secondary condensation
levels correspond to   nearly identical values of $p$.  This mechanism was
found to make possible  the decrease of mass in condensation.

\subsubsection{Calculation of cm contribution to quark mass}

The calculation of cm contribution to quark mass  differs from the
leptonic  case in some aspects. \\ a) The vacuum weights must be chosen so
that essentially same operators create  states  of given mass squared for
$\nu$ and $U$ type  quark and $ e$ and $D$ type quark.\\ b)  The
contribution of K\"ahler charge $Q_K=2/3$ of quark  to ground state
conformal  weight is

\begin{eqnarray}
\frac{Q_K^2}{2}&=&  \frac{2}{9}
\end{eqnarray}

\noindent  and  gives Planck mass for the  quark unless vacuum weight
contains
 additional term of opposite sign cancelling the contribution.   The first
 possibility is that quark confinement is related totally to color force
rather than half odd integer
 charge so  that vacuum weight contains compensating contribution

\begin{eqnarray}
h(U)&=& -2-\frac{2}{9}\nonumber\\
h(D)&=& -1-\frac{2}{9}
\end{eqnarray}

\noindent  Quarks and leptons are essentially identical apart from effects
 caused by color force.  \\ c) The second alternative to come in mind  is
based on the  rather
 attractive idea that contribution of K\"ahler charge gives free quark
Planck mass.  In hadrons $Q_K(tot) $ is integer and   light hadrons are
obtained provided that \\ i)  the contribution of K\"ahler charge is not
additive but is given by $Q_K^2(tot)/2=1/2(0)$ for baryons (mesons) and \\
ii)  the   total vacuum weight of the baryon (meson)  contains an
anomalous contribution cancelling this contribution. \\  The assumption
about additivity of vacuum weights for quarks doesn't allow this kind of
scenario.  Additivity in baryonic case implies
$h_{anom}(q)=h_{anom}(\bar{q}) =-1/6$ and in leptonic case this would give
$h_{anom}=-2/6$ and would give Planck mass for meson.   The physical
counter argument is that that quarks inside hadron are known to be
massless and in p-adic thermodynamic scnenario this is realized if free
quark is massless (the fraction of time spend in Planck mass states is
given by p-adic Boltmann factor and of the order of $1/p$!) \\ d)
Accepting the massless quark  alternative the calculation of quark masses
does not differ much form that for leptons and differences result from
differences between  N-S and Ramond type color Super Virasoro
representations.  \\ e)  To calculate hadron masses one must use
appropriate tensor product of
 Super Virasoro representations for quarks. Since quarks correspond to
different boundary components they  satisfy separately Super Virasoro gauge
conditions. This simplifies enormously the calculations  and to first
order the quark masses are additive. One might wonder whether one should
allow color excitations of quarks if they combine to form color singlet:
the results of the third paper show  that most hadrons would get  Planck
mass if this were the case. This means that free massless  quark is  the
only possible alternative.

\subsubsection{The identification of physical states and of  inner product}

Before considering the definition of the inner product it is useful to
 clarify the rules for identifying the physical states. \\ a)
$so(4)/su(2)$ and $so(3,1)/su(2)$ coset representations decompose into
 tensor products of Super Virasoro representation and Kac Moody
representation associated with  $su(2)$ degrees of freedom.  This implies
that the operators creating degenerate states can be constructed as
products of  the generators $L^n_i,G^n_i$,  where $i=1,...,6$ labels the 2
$so(3,1)$ representations , 2  $so(4)$ representations and $u(1)$ and
$su(3)$ representations appearing in the tensor product. \\
 b) The Virasoro conditions $L_n(tot)\vert phys\rangle=0$ reduce to the
four   conditions

\begin{eqnarray}
L_{n}\vert phys \rangle &=&0\nonumber\\
G_{-n+\epsilon}\vert phys \rangle&=&0 \ , n=1,2\nonumber\\
\epsilon (R)&=&0, \epsilon (N-S)=1/2
\end{eqnarray}

\noindent in ordinary case. If the representation associated with
particle   tensor products of Ramond and N-S type representations then the
conditions associated with Super generators $G^k$  separate to independent
conditions
 for N-S and Ramond representations.   \\ c)  The delicacies of the inner
product force a
 formulation of gauge conditions different from the standard formulation.
Physical states must also be orthogonal to the states of form $L^n O\vert
vac\rangle$ and Virasoro conditions can be replaced with the ortogonality
requirement.

\vm

 The definition of  a physically sensible inner product  $so(3,1)$
 and $so(4)$ degrees of freedom requires careful considerations.  In
$so(3,1)$, $so(4)$ and $su(3)$  these degrees of freedom as well in color
degrees of
 freedom
 the representations in question have $(c,h)=(0,0)$.  Formally  this
implies
 in  $so(3,1)$ and $so(4)$ degrees of freedom that
 the states created by Virasoro algebra generators  $L_n(i)$
 from vacuum  state have vanishing norm.  Only the ground state would
 correspond to nongauge degree of freedom! This is certainly not the
physical situation and the problem derives from the fact that naive inner
product is not correct. In $su(3)$ degrees of freedom zero norm for the
states created by Virasoro generators seems to be make sense.

\vm

 In p-adic case there is an elegant  manner to modify  the inner product
in $so(3,1)$ and $so(4)$ degrees of freedom.  The point is that in p-adic
case it is sensible to consider the limit $m\rightarrow 2$  by assuming
$m$ to be integer assuming that the  values of $P$ and $Q$ are just those
associated with the massless representations. One simply takes  $m$ to be
integer of form
 $m=2+ O(p^k)$ and allows $k$ to approach infinity.     In this limit the
central charge behaves as

\begin{eqnarray}
c&=& \frac{9 \Delta m}{8}\nonumber\\
\Delta m&=& O(p^k), \ k \rightarrow \infty
\end{eqnarray}

\noindent The vacuum weight $h$ behaves as

\begin{eqnarray}
h (Ramond)&=& \frac{3 \Delta m}{64}\nonumber\\
h (N-S,1,1)&=& 0\nonumber\\
h (N-S,1,3)&=& \frac{\Delta m}{4}
\end{eqnarray}

\noindent Inner products in Virasoro algebra at $m=2$ limit can be defined
 using the limiting expressions for $c$ and $h$ in the expressions for
inner products and dividing by the small parameter $\Delta m$:  just
scaling is in question.  Alternatively,  one can consider the inner
products for states,
 whose norm is taken to be equal to  one.    The inner product is unitary
for each value of $\Delta m$ in the limit and therefore unitarity  holds
true in the limit $\Delta m= 0$,  too.

\vm

The inner products involve typically commutators/anticommutators  of
various Super Virasoro generators and it is useful to list the action of
the commutators on the vacuum

\begin{eqnarray}
Comm(L^m,L^{-m})&=& 2m h  + \frac{c}{12}m(m^2-1)\nonumber\\
Anti(G^m,G^{-m})&= &2h+ \frac{c}{3}(m^2-\frac{1}{4})\nonumber\\
\
\end{eqnarray}

\noindent   The detailed study of these commutators shows that \\ a) $G^0$
creates zero norm state for Ramond representation in the  limit $\Delta m
\rightarrow 0$. \\ b) $G^{k}$ and $L^1$ create zero norm state in case of
$N-S$ $(1,1)$ representation (vanishing spin/isospin). \\ These operators
clearly act as
 super gauge symmetries. The natural physical interpretation  is as
remnant of the gauge symmetries
 defined by the entire Super Virasoro algebra.

\vm

The  crucial property of the inner product is that  for Ramond
representation
  state space decomposes into sectors according to the number $N$ of
excited $so...$  sectors and superposition for states belonging to
different sectors does not make sense since without normalization inner
products are of different order in $\Delta m$ in different sectors and
superposition with norm taken to be equal to one would imply that some
coefficients in the superposition have infinite value. Therefore one must
pose selection rule forbiding the superposition of states belonging to
different sectors. Similar phenomenon occurs in bosonic sector.

\vm

The correct manner to treat the gauge conditions is to calculate their
 consequences for $c,h\neq 0$ and to take the limit $\Delta m\rightarrow 0$
 only after that.  Furthermore,  one must consider matrix elements between
physical states and states created by $\Delta =2$ operators proportional
to Super
 Virasoro generators.  The delicacies associated with the definition of
the  inner product turn out to be important in the treatment of fermionic
gauge conditions.  It turns out that one must weaken the standard form of
gauge conditions in $N>1$  super selection sectors of state space.

\vm

Since the generators of Super Virasoro algebra  in question are generated
as
 commutators and of the generators $L^2,L^1,G^1$ one can restrict the
consideration to these gauge conditions.  This automatically eliminates
redundancy associated with gauge conditions and simplifies practical
calculations considerably.

\subsubsection{Super selection rule}

The emergence of super selection rule can be understood by studying the
general form of the neutrino state in massless case.  The general solution
to the gauge conditions can be written as

\begin{eqnarray}
O&=& O_1 +O_2 +O_3+O_5+O_c\nonumber\\
O_1&=& \sum_{i=1}^{4} ( a(i)L_i^2 +b(i) (L_i^1)^2 +c(i)  G^2_i +d(i) L^1_i
G^1_i)\nonumber\\  O_2&=&  \sum_{i}^{4}( e(i)L^1_iL^1_5  +
f(i)L^1_iG^1_5+g(i) G^1_iL^1_5+h(i) G^1_iG^1_5) \nonumber\\
O_3&=& \sum_{i\neq j} (a_{ij}L^1_iL^1_j +b_{ij}G^1_iG^1_j+
c_{ij}L^1_iG^1_j) \nonumber\\
O_5&=& eL_5^2 + f(L_5^1)^2 +g G^2_5
+hL^1_5 G^1_5\nonumber\\
O_c&=& a_c F^{3/2a}F^{1/2a}
\end{eqnarray}

 \noindent The requirement that the norm of each component in the state is
nonvanishing and finite fixes the dependence of the various coefficients
on the parameter $\Delta m$.  One has the following proportionalities:

\begin{eqnarray}
a(i),b(i),c(i),d(i) &\propto &\frac{1}{\sqrt{\Delta m}}\nonumber\\
e(i),f(i),g(i),h(i) &\propto& \frac{1}{\sqrt{\Delta m}}\nonumber\\
a_{ij}, b_{ij},c_{ij} &\propto& \frac{1}{\Delta m}\nonumber\\
a_5,b_5&\propto& 1
\end{eqnarray}

\noindent   The coefficients  become singular at the limit $\Delta m
\rightarrow 0 $ and that the dependence on $\Delta m $ is different for
the various states appearing in the decomposition ($a(i)....,h(i) \propto
1/\sqrt{\Delta m}$, $ a_{ij},b_{ij},c_{ij}\propto 1/\Delta m$.   The
states with different powers of $\Delta m$ are however  automatically
orthogonal
 and if one  assumes the superselection rule  forbiding  linear
superpositions of states belonging to different sectors there is no need
to use diverging
 coefficients if matrix elements are defined  projectively by dividing the
inner product of  two states with the norms of the states.

\subsubsection{Results for lepton and quark masses }

The results  for the leptonic degeneracies are  summarized in the following
 table

\vl

\begin{tabular}{||c|c|c|c|c|c||} \hline \hline
n& 0&1& 2 &$X_1$&$X_2$  \\ \cline{1-6}\hline
D(L)& 12&40&80&5&$\frac{2}{3}$\\ \hline
$D(\nu)$& 40&80&10&3&$\frac{1}{2}$\\ \hline\hline
D(D)& 12&40&80&5&$\frac{2}{3}$\\ \hline
D(U)& 40&80&8&3&$\frac{1}{2}$\\ \hline\hline
\end{tabular}

\vl

Table 2.1.\label{fermdeg}  The degeneracies $D(i)$, $i=01,2$ for charged
leptons
 and neutrinos and quarks.  Also are listed the coefficients
$X_1=k(F)D(3/2)/D(0)$ and $X_2\equiv ((3 (D(3)-D(3/2)^2/2D(0))\ mod \ D(0))p^2
)/D(0) $ of first and second order contributions to mass squared.

\vl

 For the choice

\begin{eqnarray}
k(F)&=&k(B)=\frac{3}{2}
\end{eqnarray}

\noindent corresponding  to mass formula $p^2= (3/2)L^0+...$  the
coefficient    $X_1=\frac{k(F)D(3/2)}{D(0)}$   of first order contribution to
 mass squared is
 integer and one has $X_1=5$ for charged leptons and $X_1=3$ for
neutrinos.  It is  rather remarkable that the condition $k(B)=3/2=k(F)$,
 which guarantees massless photon and predicts charged lepton masses with
relative error smaller than one per cent also guarantees lightness of
charged  leptons!

\vm

 For charged lepton   the coefficient of second order contribution to mass
squared is  $X_2= 2/3$.
 For Mersenne primes the real  counterpart of $2p^2/3$  is $\frac{2}{3p}$
and  same result is obtained under rather general  assumptions about $p$
associated with primary condensation level.   For the neutrino one has
  $X_2=  -1/2$.  The  real counterpart  of $-p^2/2$   depends  on the
value of $p(\nu)$:  for Mersenne prime $M_{127}$ the real counterpart  is
in good approximation $\frac{1}{2p}$.
  Similar result holds true under rather general conditions on
 $p(\nu)$.

\vm

The coefficients of first order contributions to mass squared are
identical for
 leptons and quarks.
  The  difference  in second order contribution of $U$ and $\nu$  results
 from the difference between Ramond and N-S type color Super Virasoro
algebras.
 The over all mass scale for quark masses is determined by  the
condensation
 level, which
 must correspond to prime  $k=107$ for $u,d,s$ and $k=103$ for $c$ and $b$
as  will be found in the third part of the paper.

\vm

Summarizing,   under rather general conditions on $p$ associated with the
primary condensation level  the expression for the cm contribution to
lepton mass and its real counterpart reads as

\begin{eqnarray}
M^2(cm,L)&=&(5p+ \frac{2}{3}p^2)M_0^2 \nonumber\\
M^2(cm,L)_R&=& (5+\frac{2}{3})\frac{M_0^2}{p(L)}\nonumber\\
M^2_R(cm,\nu) &=&
(3p+ \frac{27}{20} p^2)M^2_0\nonumber\\
M^2_R(cm,\nu)_R &=&
(3+ \frac{7}{10}) \frac{M^2_0}{p(\nu_e)}
\end{eqnarray}

\noindent For electron no modular contribution is present so that a
prediction
 for electron gauge boson mass ratio follows:

\begin{eqnarray}
\frac{M_W}{m_e}&=& \sqrt{\frac{M_{127}}{M_{89}}}
 \sqrt{\frac{1}{2(5+\frac{2}{3})}}
\end{eqnarray}

\noindent    $W$ mass ($m_W\simeq 80.8 \ GeV$ experimentally)
 is predicted to be too small by $0.7$   per cent.    $Z^0$ mass is
predicted too large by $10$  per cent and th error derives the too large
value of Weinberg angle ($sin^2(\theta_W)=3/8$).  Topological condensation
is expected to renormalize Weinberg angle by reducing $Z^0$ mass.

\vm

There are also some exotic light  fermions.  The calculations of the
appendix show that  all leptons allow color decuplets
 ($10$ and $\bar{10}$)
 and neutrinos also twice degenerate $27$-plet as
 massless state.
  $U$ type  quarks allow also  massless color decuplets.
 The masses of color  excitations given in the table below.

\vl

\begin{tabular} {||l|l||}\hline \hline
fermion  &$M_R/m_e\sqrt{\frac{M_{127}}{p}}$ \\
\hline  $L^{10}$  & $\sqrt{\frac{9}{5+\frac{2}{3}}}$  \\
\hline
$\nu_L^{10},\nu_L^{\bar{10}}$ &$
1 $\\ \hline
$\nu_L^{27}$ &  $\sqrt{\frac{9}{5+\frac{2}{3}}}$    \\
\hline
 $U^{10},U^{\bar{10}}$&  $\sqrt{\frac{9}{5+\frac{2}{3}}}
$ \\  \hline\hline
\end{tabular}

\vl

Table 2.2. \label{Exoticmasses}  The masses of color excited  leptons and
quarks.

\vm

\noindent
The existence of colored electron implies new branch of physics unless the
primary condensation level for some reason corresponds to  small $p$ rather
 than $M_{127}$. Recall that the  TGD  inspired explanation for anomalies
$e^+e^-$ pairs observed
 in heavy ion collision \cite{Heavy,Lepto} is in terms of leptopions,
which are color bound  states of color excited leptons.

\subsection{Calculation of  bosonic  masses}

Elementary boson mass squared is assumed to be given by boundary
contribution,
 which can be expressed as sum of cm and modular contributions:
$M^2=M^2(cm) +M^2(mod)$.  The fact that higher boson families  are not
observed does not necessary implies Planck mass for higher boson
families.  The construction of elementary particle vacuum functionals
\cite{TGD} demonstrated that the amplitudes for transitions such as the
decay of muon to electron by emission of $g=1$ boson vanish for vacuum
functionals.  If $g>0$ families are light  their mass is at least of the
order of intermediate gauge boson mass if one takes seriously the
observation that the absence of tachyons for bosonic Dirac operator leaves
only $M_{n}$,$n=89,61,17$.

\vm

 The construction of the cm contribution $M^2(cm)$ to  bosonic mass relies
on the following  physical  picture.\\ a) Bosons are described using
generalized spinors and obey
 generalized Dirac equation leading to mass shell condition
$p^2=k(B)L^0+$.  As already found the absence of actual  tachyons is
guaranteed if one has $k(B)=k(F)=3/2$. Same result follows also from the
requirement that photon is massless.\\ b)  Operators with half odd integer
conformal weight transform leptons to  quarks
 and vice versa.  Leptoquarks, that is bosons transforming leptons to
baryons and vice versa,  are not possible
 at classical level and G-parity rule motivated by  tachyon elimination
excludes
 these states.  The rule also guarantees that partition function contains
only integer powers of $p$ (rather than half integer powers). \\ c)  The
temperature parameter $ T=1/n$, $n=1,2,3...$ appearing in the partition
function is a free parameter at this stage, which could even depend on
particle.  It turns out
 that $T(ew)=1/2$ is necessary to understand the masses of electroweak
gauge bosons. $T=1$ provides a manner to eliminate  exotic massless states
from spectrum. As a consequence,   the contribution to the  thermal masses
from  $\Delta =1$ level
 is of order $O(1/p^2)$ for gauge bosons.  The condition $D(3/2) \ mod \
D(0)=0$ guarantees that boson  is essentially massless.

\vm

Consider next the technical aspects of the state  construction.\\ a)
Boson state must contain information about its couplings to fermions.
 It is useful to separate Super Virasoro and Kac Moody degrees of freedom,
where charge matrices act. Bosonic charge matrices are linear combinations
of $CP_2$ sigma matrices and K\"ahler charge times unit  matrix.
 The Kac Moody counterparts for vectorial and axial  isospins and K\"ahler
charge are the matrices $F^{1/2,3}_3$,  $F^{1/2,3}_4$  and $F^{1/2}_5$.
The counterparts of $W$ boson charge matrices are linear combinations of
the matrices $F^{1/2,i}_3F^{1/2,j}_4$, $i,j=1,2$.  For gluons the charge
matrices are the operators $F^{A1/2}$, $A=1,..,8$.   Bosons are therefore
generated from ground states of form  $\vert phys \rangle=
EQ^k_B\vert vac\rangle$, $  E= \epsilon_k\gamma^k_{1/2}$.
  For spin one bosons the  state is also proportional
 to the operator $E$ of conformal weight $1/2$ defined by the
polarization vector of gauge boson.  The state satisfies automatically
gauge conditions.   Massless states are obtained by applying operators of
conformal weight
 $\Delta=  2-k$  to the ground state.  The generalized Dirac equation
implies the conditions $p^2=0$ and $p \cdot \epsilon=0$ for the  massless
states.\\ b) It  is not clear whether one should allow all isospins for the
operators $F^{1/2,i}_k$ (isospin index has not been written in the
formulas of appendix). The fact that intermediate gauge bosons correspond to
subset of all possible index combinations suggests that there is some
hitherto
unidentified condition  excluding some isospins. This kind of rule might  be
the analogy of Dirac equation  in $so(3,1)$ degrees of freedom and would mean
the introduction of $so(4)$  «momentum«  and «polarization«.   $so(4)$
polarization could have something to do with the direction defined by the
vacuum expectation value of Higgs field in standard model. Unfortunately,
the situation is unclear at this stage.\\
 c)  For intermediate gauge  bosons also longitudinal
polarization  operator
 $P=p_k\gamma_k^{1/2}$  is present  in massless sector since otherwise
longitudinal polarization would correspond to Planck mass excitation. It is
necessary to define the norm of longitudinally polarized massless state by
dividing with $\sqrt{p^2}$ and taking the limit $p^2\rightarrow 0$.  This
 indeed makes sense if one defines inner product in $so(4) $ degrees of
freedom  as the p-adic limit
 $m=2+\Delta m\rightarrow 2$ so that $p^2\propto h \propto \Delta m$
holds true. In the  actual physical situation a small value of $\Delta m$
might well be  generated by secondary
 topological  condensation so that limiting procedure could be regarded as
a useful mathematical  idealization.   The counterpart of this in ordinary
description of massivation is the transformation  of the gradient of Higgs
field to the longitudinal polarization of gauge boson.  Since $P$ is not
 used in state construction for transversally  polarized states there is
complete symmetry
 between transversal and longitudinal polarizations and the calculations
performed for transversal  polarization apply as such for longitudinal
polarization.
 One can choose the coordinates so that polarization vector
 corresponds to $i=2$ tensor factor and $P=p_k\gamma^k_{1/2}$ corresponds
to $i=1$. \\ d)  The operators creating massive excitations are just the
Super Virasoro  generators associated with different tensor factors in
$so...$ degrees of freedom.   In color degrees of freedom the
multiplication with Super Virasoro generators produces zero norm states
and  one must use commutator action instead.  An alternative possibility
is that no excitations are allowed in color degrees of freedom. In color
degrees of freedom there is possibility of forming nonlinear combinations
of Kac Moody generators, which would imply nonlinear terms in couplings to
fermions.  It turns out that  massless gluon is obtained   if only linear
combinations are allowed.  \\ e)  In fermionic case  it was necessary to
modify the inner product for
 Super Virasoro, which led to the decomposition of the state space to super
selection sectors  labeled by  $N=0,1,2,.$.  Also in bosonic case one can
consider the possibility of including the states  created by $L^n$ $n\ge
2$   and $G^k, k\ge 3/2$  from N-S singlet vacuums in $so...$ degrees of
freedom.  For ordinary inner product these states possess vanishing norm
but one could modify the norm in the same manner as in fermionic case and
obtain additional super selection sectors to the state space.  The
construction of boson states  serves as test for this alternative.   It
has not been possible to identify any working scenario allowig the
inclusion of $N\ge 0$ sectors of state space.  It should be noticed  that
for $u(1)$ sector the states created by polynomials of $L^k,k\ge 0$ and
$G^k,k\ge 3/2$ possess nonvanishing norm  as is clear from the commutator
algebra ($G^{3/2}_5$ creates state with nonvanishing norm!) and must be
taken into account in state construction.

\vm

The results are listed in the tables and will be   discussed in detail in
subsequent paper.
    The masses of gauge bosons are in
 quantitative accordance with expectations assuming $T(ew)=1/2$ whereas
 $T=1$ must be assumed for exotic bosons. The remarkable result  is the
absence
 of   massless exotic bosons.

\vl

\begin{tabular}{||c|c|c|c||}\hline\hline
boson&$M_{obs}/MeV$& $M_{pred}/MeV$&error/\%\\ \hline \hline
$\gamma$&0&0&0\\ \hline\hline
gluon&0&0&0 \\ \hline\hline
 $W$ &80200&79582&-0.8\\ \hline\hline
$Z$ &91151&100664&10.0\\ \hline\hline
\end{tabular}

\vl

Table 2.3.  \label{Bosonmasses}  Masses of nonexotic gauge bosons.

\vl

\begin{tabular}{||c|c|c|c||}
 \hline\hline
spin  &charge operator &$ M^2 (T=1)$ &$ M^2 (T=1/2)$  \\ \cline{1-4}\hline
0&$I^3_{L/R}$& Planck mass&$\frac{1}{2p}$\\ \hline
0&$I^{\pm}$ & Planck mass &$\frac{1}{2p}$  \\ \hline
0&$I^3_{L/R}Q_K$ &$\frac{3}{p}$&0\\ \hline
0&$I^{\pm}Q_K$ &Planck mass&$\frac{1}{2p}$\\ \hline
1&$1$ &Planck mass&$\frac{1}{2p}$\\ \hline\hline
\end{tabular}

\vl

Table 2.4. \label{Exobosonmasses} Masses and couplings of noncolored light
 exotic
bosons for $T=1$  and $T=1/2$.  Charge operator
 tells how the boson in question couples to matter. For $T=1/2$
the states with charge operator $I^3_{L/R}Q_K$ are essentially
massless for large values of $p$ and some additional light states
 become possible.

\vl

\begin{tabular}{||c|c|c|c|c||}
 \hline\hline
spin  &charge operator & $D$ &$ M^2 (T=1)$&$M^2(T=1/2)$ \\ \cline{1-5}\hline
0& $I^{\pm}Q_K$ &8 &$\frac{2}{p}$&0\\ \hline
1& $I^{\pm}$ &8 &Planck mass&$\frac{1}{2p}$\\ \hline
1& $I^3_{L/R}Q_K$ &8 &Planck mass&$\frac{1}{2p}$\\ \hline
1&$I^{\pm}Q_K$&8 &Planck mass &$(\frac{3p}{10})_R\frac{1}{p}$
\\ \hline\hline
0& $I^{\pm}Q_K$ &$10,\bar{10}$ &$\frac{3}{p}$&0\\ \hline
0& $1$ &$10,\bar{10}$ &0&$0$\\ \hline
1&$I^{\pm}$&$10,\bar{10}$ &$\frac{3}{p}$&0\\ \hline
1&$I^3_{R/L}Q_K$&$10,\bar{10}$ &$\frac{3}{p}$&0\\ \hline\hline
0&$I^{\pm}$&27 &Planck mass &$\frac{1}{2p}$\\ \hline
0&$I^3_{R/L}Q_K$&27 &Planck mass &$\frac{1}{2p}$\\ \hline
1&$I^3_{R/L}$&27 &Planck mass &$\frac{1}{2p}$\\ \hline
1&$Q_K$&27 &Planck mass &$\frac{1}{2p}$\\ \hline\hline
0&$I^3_{R/L}$&27 &0 &0\\ \hline
0&$Q_K$&27 &0 &0\\ \hline
1&$1$&27 &0 &0\\ \hline\hline
\end{tabular}

\vl

Table 2.5. \label{Colbosonmasses} Masses and couplings of colored
 exotic
bosons for $T=1$ and $T=1/2$.  The last two massless bosons are doubly
 degenerate due to occurrence of two $27$-plets with conformal weight
$n=2$. $T=1/2$  is physically possible alternative since no long range
 forces
are implied.

\vl

\begin{center}
{\bf Acknowledgements\/}
\end{center}

\vm

It would not been possible to carry out this work without the  concrete
help of
 my friends  in concrete problems of the everyday life and I want to
express my gratitude to  them.  Also I want to thank J. Arponen,
 R. Kinnunen and J.  Maalampi
    for practical help and interesting discussions.

\newpage

\section{ Appendix A:Calculation of degeneracies   for neutrinos and U type
 quarks}

In the following the thermal expectation for the cm contribution to mass
squared
 of neutrino  and U quark is
 derived.    The  differences between  neutrino and U quark  derive from
the
 differences between Ramond (quarks) and N-S type (leptons)
representations of Super Virasoro in color sector. The first difference
comes in order $O(p^2)$.  The  results of calculation can be used in the
considerably simpler calculation of $M^2(cm)$ for charged leptons and D
quarks.

\vm

 Supersymmetry requirement leads to some delicate considerations for $L^2$
 gauge conditions necessitated by the limiting procedure appearing in the
definition of the inner product.  States have well defined number of
$so(4)$  type supergenerators $G^k_i$ , which is odd or even.
Supersymmetry means  that the solutions with odd number of super
generators are obtained by the applicatication of $G^0$ from the solutions
with even G-parity.  This implies that in certain cases the gauge
conditions for $L^2$ implies  two conditions instead of one condition as
one might expect. The point is that $L^0_i$ terms in state $L^2X$ give
scalar term proportional to $h (\rightarrow 0)$ and the coefficient of
this term must vanish. Also terms proportional to $G^0_i$
 appear in the state $L^2X$ and  give zero norm state.  The application of
$G^0$ to the state $L^2X$ however transforms these terms to terms
proportional to $L^0_i$ and implies doubling of  $L^2$ gauge condition.
The resulting gauge condition is just that for $G^2$ and would follow for
ordinary inner product as a consequence of $L^1$ and $G^1$ gauge
conditions.

\vm

The following tables summarize the results for the degeneracies for various
 mass squared values  in various  super selection sectors.

\vl

\begin{tabular}{|| c|c|c|c|c|c|c||}\hline \hline
$M^2$& N=0&N=1&N=2&N=3&N=4&D\\ \cline{1-7}\hline
0& 0&20&20&0&0&40\\ \hline
$\frac{3}{2}$&0&44&32&4&0&80\\ \hline
3&2&8&0&0&0&10\\ \hline\hline
\end{tabular}

\vl

Table 3.1.\label{neutridege} The degeneracies of  neutrino states with
$M2=0,3/2,3$  in super selection sectors with $N=0,1,2,3,4$.  Last column
gives total degeneracies.

\vl

\begin{tabular}{|| c|c|c|c|c|c|c||}\hline \hline
$M^2$& N=0&N=1&N=2&N=3&N=4&D\\ \cline{1-7}\hline
0& 0&20&20&0&0&40\\ \hline
$\frac{3}{2}$&0&44&32&4&0&80\\ \hline
3&0&8&0&0&0&8\\ \hline\hline
\end{tabular}

\vl

Table 3.2.\label{Uquarkdegen} The degeneracies of  U quark  states with
 $M^2=0,3/2,3$  in super selection sectors with $N=0,1,2,3,4$.  Last
column gives total degeneracies.

\subsection{Degeneracy of $M^2=0$ states of U quark and neutrino}

Super selection rule implies that gauge conditions can be applied
separately  in each sector.

\vm

{\bf 1.  $N=0$ sector. }

\vm

For  $N=0$ states the gauge conditions correspond to
  $4$ operators $O^2_5$ ($4$) and the operator  $O^2_c=
\sum_aF^{a3/2}F^{a1/2}$
 acting in color degrees of freedom: $5$ altogether.  This operator is
 present also in quark sector.  Gauge conditions for $L^1$ and $G^1$
correspond to operators $O^1_5$($2$)  and  gauge condition for $L^2$
corresponds to a multiple of unit operator ($1$). The number of gauge
conditions is $2+2+1=5$ so that no solutions to gauge conditions are
obtained: $D(0,0)=0$.

\vm

{\bf 2. $N=1$ sector}

\vm

The $32$ states in   $N=1$ sector are created \\
i) by $16$ single particles operators  $O^2_i $ given by
 $L_i^2$, $(L_i^1)^2$, $ G^2_i$, $L^1_i G^1_i$.\\
ii)  by  $16$ 2-particle operators $O^{1,1}_{i5}$
with $O_{i5}$ given by   $ L^1_iL^1_5 $, $ L^1_iG^1_5$, $
G^1_iL^1_5$, $ G^1_iG^1_5$, $i\leq 4$   coupling  $so(3,1)\times so(4)$
and $u(1)$ degrees of freedom to each other.

\vm

Gauge conditions for $L^1$ and $G^1$ correspond to operators \\
 $O^1_i$,$i=1,...,5$,  ($5\cdot 2=10$)  and gauge conditions for $L^2$
correspond to unit matrix.   The   orthogonality with respect to states
created by the operators   $L^2$ gives actually two  conditions since
$G^0$ acts as super symmetry and states can be classified in two types
with general forms

\begin{eqnarray}
O&=&\sum_i (a_i L^2_i + b_i (L^2_i)^2 + c_i L^1_iL^1_5+d_iL^1_i G^1_5
)\nonumber\\
O&=&\sum_i (a_iG^2_i + b_i L^1_iG^1_i + c_i G^1_iL^1_5+d_iG^1_i G^1_5
)\nonumber\\
 \
\end{eqnarray}

\noindent  and the condition comes from
 the coefficient of $L^0$ term in commutator.    The number of conditions
is
 therefore $22$.  This gives $32-22=10$ solutions to gauge conditions and
taking into account the doublefold degeneracy associated with $U(1)$
degrees of freedom one has $D(0,1)=20$.

 \vm

{\bf 3. $N=2$ sector.}

\vm

 There are $24$  two-particle operators $O^{1,1}_{ij}$, $i \neq j$
where $O^{1,1}_{ij}$ is  given by $ L^1_iL^1_j$, $ L^1_iG^1_j$, $
G^1_iL^1_j$, $ G^1_iG^1_j$, $i\neq j\leq 4$.
  The application of gauge
conditions for $L^1$ and $G^1$  gives

\begin{eqnarray}
\sum_{j}a_{ij}&=&\sum_j
b_{ij}= \sum_j c_{ij}=\sum_jc_{ji}=0
\end{eqnarray}

\noindent   and the number of operators is reduced to $10$.   Naiive
counting of gauge conditions would give $8+8=16$  gauge conditions
corresponding to operators $O^1_i$ associated with $L^1$ and $G^1$ gauge
conditions.  The reason for redundancy is that for the antisymmetric part
of the coefficient matrix the gauge conditions are redundant. To sum up,
one has $D(0,2)=20$. and $D(0)=40$.  There are no differences between U
quark and neutrino.

\subsection{Degeneracy of  $M^2=3/2$ states for neutrino and U type quark}

$M^2=3/2$ states are created by the operators of conformal
 weight $\Delta =3$ from a Ramond ground state with definite quantum
numbers.
 $M^2=3/2$  operators can be labeled by the number  $N$ defined as the
number
 of $so...$ type Super Virasoro generators appearing in them and linear
super position for states with different $N$ is forbidden  by super
selection rule.  $N$ can have the values $N=0,1,2,3$ for $M^2=1$ states.
It is useful to introduce some shorthand notations. $O^n_i$ refers to
 $\Delta =n$ operators in sector $i=1,...,5$.  $O^1_i$ refers to
 $L^1_i$ and $G^1_i$.  $ O^2_i$ refers  to  the $4$ operators
$L^2_i,G^2_i$,
 $(L^1_i)^2, L^1_iG^1_i$.  $O^3_i$ refers to the $8$ operators
 $L^3_i, G^3_i$ $,L^2_iL^1_i, (L^1_i)^3$, $ G^2_iG^1_i,L^2_iG^1_i$,
  $(L^1_i)^2G^1_i, L^1_iG^2_i$.

\vm

{\bf1. $N=0$ sector.\/}

\vm

$N=0$ operators correspond to operators acting in $u(1)\times su(3)$
degrees
 of freedom only.  There are following operators. \\ a) The $8$ operators
$O^3_5$   acting in $u(1)$ degrees of freedom.\\   b) Color singlet
operators $O^3_c(2) $ satisfying gauge conditions. From the  table
\ref{Nsginvdege} of appendix F  one finds that there are no color singlet
operators  satisfying gauge conditions.  \\ c) The operators $
O^1_5F^{a3/2}F^{a1/2}$($2$). \\ Total number of operators is  $8+2=10$.

\vm

 The total  number of gauge conditions is $4+4+2=10$ and is same larger
than the number of operators  so that one has  $D(1,0)=0$.  Same
calculation applies to U quark since also in this case there is one  color
singlet
 operator $O^2_c$ satisfying all gauge conditions except the gauge
conditions
 associated with $L^2$ (see table \ref{Raginvdege}.

\vm

{\bf 2. $N=1$ sector.\/}

\vm

In $N=1$ sector there are following operators. \\ a) The  $32$ operators
$O^3_i$, $i=1,...,4$   acting in $so... $ degrees
 of freedom.  \\ b) The $32$ operators $O^2_iO^1_5$   and the $32$
operators of of form  $O^1_i O^2_5$.\\ c) The $8$ operators $O^1_i
F^{a3/2}F^{a1/2}$ , $i=1,...,4$ acting in
 $so...\times su(3)$ degrees of freedom. These operators satisfy all gauge
conditions associated with $G^k$, $k=1/2,..,5/2$. The only nontrivial gauge
conditions are associated with  $L^1$ and $L^2$.\\ The total number of
$N=1$ operators is  $104$.

\vm

\noindent a) Gauge conditions for $L^1$ and $G^1$  correspond to  i)  the
$32$  operators  $O^2$ in $N=1$ sector. There is  a rather delicate
reduction of gauge conditions associated with $G^1$: these  gauge
conditions are equivalent with the requirement that physical states are
orthogonal to states of form $G^1O^2$ and for the states $O^2=
G^1_iG^1_5$, $i=1,..,4$ the ortogonality condition is identically
satisfied since the projection of $O^2$ to $N=1$ sector vanishes
automatically by anticommutativity of $G^1_i$. This implies the reduction
of gauge conditions by $4$ to $28$.\\ ii) the  $5$ operators  $O^2$ in
$N=0$ sector.  These operators were already listed, when evaluating  the
degeneracy of massless states. \\ Altogether thre are  $32+28+10= 70$
conditions. \\ b) Gauge conditions for $L^2$ correspond to the $8$
operators $O^1_i$ and $2$ operators $O^1_5$. The requirement of
supersymmetry doubles the gauge conditions associated with $O^1_5$ so that
the number of conditions is actually $8+4=12$.   \\ The total number of
gauge conditions is $32+28+10+12= 82$ and  the number of physical
operators is $104-82=22$. The contribution to degeneracy is $D(1,1)=44$.

\vm

{\bf 3. $N=2$ sector. }

\vm

There are following $N=2$ operators.\\ a) The operators $O^2_i O^1_j$, $i
\neq j =1,...,4$, whose total number  is $ 96$. \\ b)  The operators
$O^1_i O^1_j O^1_5$, $ i \neq j =1,...,4$, whose total number is $48$.\\
The total number of operators is $96+48= 144$.

\vm

\noindent a) Gauge conditions for $L^1$ and $G^1$ correspond   to \\ i)
$24$  operators $O^1_iO^1_j$   in $N=2$ sector. \\ ii) $ 16 $ operators
$O^1_iO^1_5$  and  $16$ operators $O^2_i$ in $N=1$ sector:
 altogether $32$ operators.\\ The total number of conditions is $2 \cdot
56=112$.\\ b) Gauge conditions for $L^2$ correspond to $8$  $N=1$ operators
$O^1_i$.  The number of conditions get doubled by super symmetry.\\
  This means that the number of conditions becomes $112+16=128$ and the
number
 of states becomes $144-128=16$ and the contribution to  the degeneracy is
$D(1,2)=32$.

\vm

{\bf 4. $N=3$ sector. }

\vm

$32$  operators $O^1_iO^1_jO^1_k$, $i\neq j\neq k=1,..,4$
 contribute to the degeneracy in $N=3$ sector.    For the states
 $a_{ijk}G^1_iG^1_kG^1_k$ and  $b_{ijk}G^1_iG^j G^1_k$ gauge conditions
 allow one solution for each and the the contribution to degeneray  is
$D(1,3)=4$. The total degeneracy of $M^2=3/2$ states is $D(3/2)=80$. Same
result is obtained for $U$ quark.

\subsection{Degeneracy of   $M^2=3$ states for neutrino and U quark}

The degeneracies of fermionic $M^2=3$ states is evaluated for $N=0,1,2,3,4$
 sectors.   The difference between U quark  and neutrino  emerges first
time for
 $M^2=3$ states.

\vm

{\bf 1.  $N=0$ sector}

\vm

 The operators creating $N=0$  states  are $O^4_5$($14$),  $O^4_c(6) $,
$O^3_cO^1_5(4)$
 and $O^2_cO^2_5$($4$):$14+6+4+4= 28$ altogether.  The numbers of the
color singlet
 operators  $O^k_c$,  which satisfy gauge conditions for $L^1,G^{1/2}$ and
possibly for $L^2$  can be easily found from the table of  appendix F
 (table \ref{Nsginvdege}).

\vm
Consider next gauge conditions. \\ a) Gauge conditions for $L^1,G^1,L^2$
imply that the coefficients of operators $O^4_5$, $O^3_cO^1_5$ and
$O^2_cO^2_5$ vanish. The gauge conditions for $O^4_c$
 leave
 only single operator $O^4_c$ so that  the contribution to degeneracy is
$D(2,0)=2$.

\vm

For U quark there exist no gauge invariant  color singlet  operators
$O^4_c$ as
 the table of appendix F shows so that $D(2,0)=0$ for $U$ quark.  There
are no further
 differences between U quark and neutrino.

\vm

{\bf 2.  $N=1$ sector}

\vm

The following operators are present in $N=1$ sector.\\ a) Single particle
operators $O^{4}_i$ ($13\cdot 4=52$).\\ b) Two particle operators
$O^3_iO^1_5$($8\cdot 2\cdot 4=4\cdot 16$),
 $O^2_iO^2_5$ ($4\cdot 4 \cdot 4=64$).  $O^1_i O^3_5$ ($2\cdot 8\cdot
4=64$): $4\cdot 48$ altogether.\\ c) $O^3_cO^1_i$(16) ( these operators are
 eliminated by gauge conditions for $G^{1/2}$ and $L^1$), $O^2_cO^2_i$
 ($4\cdot 4=16$), $O^2_cO^1_5O^1_i $ ($2\cdot 2 \cdot 4=16$).  The
operators $O^k_c$ satisfy the gauge conditions in color degrees of freedom
except possibly the gauge condition for $L^2$. There are $4\cdot 8$
operators altogether\\
 The total number of operators is  $4\cdot (13+48+8)=276$.\\

\vm

Gauge conditions for $L^1$ and $G^1$ correspond to  $N=1$  operators
 $O^3_i$($104$) and $N=0$ operators  $O^3$($11$): $104+11=115$
altogether.  Gauge conditions for $L^2$ correspond to $N=1$ operators
$O^2_i$ ($32$) and $N=0$  operators $O^2_5$($4$) and $O^2_c$($1$): $37$
altogether. Supersymmetry implies  doubling for of $L^2$ gauge conditions
in $N=0$ sector and one has $42$ conditions.  The total number of gauge
conditions is  $115+115+42= 272$ and the total number of allowed operators
is $276-272=4$ and one has $D(2,1)=8$

\vm

{\bf 3.  $N=2$ sector}

\vm

The following operators are present in $N=2$ sector.\\ a) The   operators
$O^{2}_iO^2_j$ ($6\cdot 4\cdot 4=6\cdot 16$).\\ b) The operators
$O^3_iO^1_j$($12\cdot 8\cdot 2= 12\cdot 16$)\\ c) The operators
$O^2_iO^1_j O^1_5$ ($12\cdot 4 \cdot 2\cdot 2= 12\cdot 16$)\\ d)  The
operators $O^1_iO^1_jO^2_5$ ($6\cdot 2 \cdot 2 \cdot 4 =6\cdot 16$)\\ e)
$O^1_iO^1_j O^2_c$ ($6\cdot 2\cdot 2 =24$)\\ The total number of states
is  $12\cdot 35=426$.

\vm

Gauge conditions for $L^1$ and $G^1$ correspond to $N=2$ operators
 $O^3_{ij}$($144$) and $N=1$ operators $O^3_i$($104$). Gauge conditions for
$L^2$ correspond to operators $O^{2}_{ij}$( $24$) and $N=1$ operators
$O^2_i$($32$): super symmetry doubles the gauge conditions associated with
$O^2(i)$. The total number of gauge conditions is $144+144+104+104+ 24+
64=584$ , which
 exceeds the number of the  operators at hand  so that the contribution to
degeneracy is $D(2,2)= 0 $.

\vm

{\bf 4.   $N=3$ and $N=4$ sectors }

\vm

The operators creating states in $N=3$ sector are:\\
a) $O^2_i O^1_jO^1_k$ ($4\cdot 3\cdot 4\cdot 2 \cdot 2=12\cdot 16$)\\
b) $O^1_iO^1_jO^1_kO^1_5$ ($4 \cdot 16=64$).  \\
The total number of operators is $ 16^2$.

\vm

Gauge conditions for $L^1$ and $G^1$ correspond to the operators\\ i)
$N=3$ operators $O^1_iO^1_jO^1_k$ ($4\cdot 8=32$).\\ ii) $N=2$ operators
$O^2_iO^1_j$ ($12 \cdot 4 \cdot 2=8\cdot 12$) and $O^1_iO^1_jO^1_5$( $12
\cdot 2\cdot 2+ 6\cdot 2\cdot 2= 6\cdot 12$). The total number of
conditions is $14\cdot 12+ 32$.\\ Gauge conditions for $L^2$ correspond to
operators $O^1_iO^1_k$  ($12 \cdot 2+6\cdot 2= 3\cdot 12$).  Total number
of gauge conditions is  $31 \cdot 12+64$ and larger than the number of
operators  ($16^2$) so that
 no solutions to gauge conditions are obtained.  Situation is same in
$N=4$ sector and one has $D(2,3)=D(2,4)=0$. The total  degeneracy for
neutrino  is $D(2)=10$.  For $U$  quark the total degeneacy is $D(2)=8$.

\section{Appendix B:  Degeneracies  for charged leptons and D type quarks}

The calculation of $M^2(cm)$ for charged leptons is easy task since  the
operators creating $M^2=0$ and $M^2=3/2$ states have conformal weights
 $\Delta=2$ and $\Delta=1$.   The operators $O^1$ creating massless
states  are linear combinations of
  $8$ operators $O^1_i$ and gauge conditions   for $L^1$ and $G^1$ reduce
their number to $6$, which gives $D(0)=12$.  From the  calculation of
neutrino
 masses  one has $D(3/2)=40$ and  $D(3)=80$.  The table summarizes the
results.

\vl

\begin{tabular}{|| c|c|c|c|c|c|c||}\hline \hline
$M^2$& N=0&N=1&N=2&N=3&N=4&D\\ \cline{1-7}\hline
0&     0&1&   0& 0& 0&12\\ \hline
3/2& 0&20&20&0&0&40\\ \hline
3&0&44&32&4&0&80\\ \hline\hline
\end{tabular}

\vl

Table 4.1.\label{Ldege} The degeneracies of  charged leptons and $D$ type
quarks  with $M2=0,3/2,3$  in super selection  sectors with
$N=0,1,2,3,4$.  Last column gives total
 degeneracies.

\section{ Appendix C: Degeneracies for colored excitations of leptons
and quarks }

Before going to the results  it is useful to notice
 that color super generators $F^{A1/2}$ are leptoquarks by the previous
results. This means that the states created by operators with half  odd
integer $\Delta$ are in fact carriers of baryon number, that is  quarks!
This in turn implies that quarks would belong to triality zero
representation and this is not in accordance with Quantum TGD.  One manner
to get out of difficulties is to assume strong form of G-parity rule: only
$\Delta \in Z$ excitations are allowed and leptoquarks are not possible.
One can also consider the possibility that for many lepton states only the
entire state satisfies G-parity rule but composite leptons are allowed to
exist in states of wrong G-parity.

\vm

The calculation of degeneracies for colored excitations reduces to the
previous  calculations for ordinary leptons:   \\ a) The table
\ref{Nsginvdege} of appendix F about color Super Virasoro shows  that only
gauge invariant colored representations with conformal weight not larger
than  $2$  are $10,\bar{10}$ with conformal weight $n=1$ and $27$ with
conformal weight $n=2$.  This means
 that  all leptons allow color decuplet massless state and neutrinos also
27 dimensional massless state.  \\ b)  The degeneracies of corresponding
states are easily evaluated.  For decuplet the   operators creating states
with given mass squared have
 conformal weight one unit smaller than for ordinary leptons.  For
decuplet these operators have conformal weight two units smaller than for
ordinary neutrino. \\ c) Also $U$ type  quarks allow massless color
excitations as the table for
 multiplicities for gauge invariant color multiplets in Ramond
representation shows. The
 excitations correspond to $10,\bar{10}$ and have conformal weight $n=2$
so that are possible for $U$ quarks only. \\ d) The results imply that the
previous  calculations for leptons and quarks make
 possible to deduce the degeneracies of color  excitations given in the
table below.

\vl

\begin{tabular} {||c|c|c|c|c||}\hline \hline
fermion & D(0)        & D(3/2)& D(3) &$M_R/m_e\sqrt{\frac{M_{127}}{p}}$ \\
\hline  $e^{10}$ &2  & 12  & 40 & $\sqrt{\frac{9}{5+\frac{2}{3}}}$  \\
\hline
$\nu^{10}$ & 12& 40 & 80& $
1 $\\ \hline
$\nu^{27}$ & 2& 12 & 40&   $\sqrt{\frac{9}{5+\frac{2}{3}}}$    \\
\hline
 $U^{27}$& 2& 12&40& $\sqrt{\frac{9}{5+\frac{2}{3}}}
$ \\  \hline\hline
\end{tabular}

\vl

Table 6.1.\label{Fcolordege}  Degeneracies and masses for light  color
excitations  for fermions. Note that $27$-plet appears twice.

\section{ Appendix D: Detailed calculation of gauge boson masses}

The results for  bosonic degeneracies for nonexotic  gauge bosons are
summarized in the following table.

\vl

\begin{tabular}{|| c|c|c||}\hline \hline
Boson& $M^2=0$&$M^2=3/2$\\ \cline{1-3}\hline
$\gamma$&3&   2  \\ \hline
$Z^0$&5&4\\ \hline
$W^{\pm}$&3&11\\ \hline
\end{tabular}

\vl

Table 6.1.\label{Bosodege} The degeneracies of $M^2=0$ and $M^2=3/2$ states
for
 electroweak gauge bosons.

\vm

The degeneracies of exotic bosons are listed in table below.  The charge
 operator $O$  associated with ground state characterizes the boson in
question.

\vl

\begin{tabular}{|| c|c|c|c|c||}\hline \hline
Type&$O$& $M^2=0$&$M^2=3/2$&$M^2$\\ \cline{1-5}\hline
$J=0$&1&0&     &ah\\ \hline
&$F_5^{1/2}$&0&&ah\\ \hline
&$F_k^{1/2},k=3,4$&1&1&h\\ \hline
&$F_3^{1/2}F^{1/2}_4$&2&3& h\\ \hline
&$F_k^{1/2}F^{1/2}_5$,$k=3,4$&1&2&$\frac{3}{p}$\\ \hline
&$F_3^{1/2}F^{1/2}_4F^{1/2}_5$&2&10&h\\ \hline\hline
$(J=0,8)$ &$1$&0&&ah \\ \hline
&$F^{1/2}_k,k=3,4$&0&&ah\\ \hline
&$F^{1/2}_3F^{1/2}_4$&0&&ah\\ \hline
&$F^{1/2}_3F^{1/2}_4F^{1/2}_5$&3&4&$\frac{2}{p}$\\ \hline\hline
$J=1$ &$E=p^k\gamma_k^{1/2}$&1&1&h
\\ \hline
$(J=1,8)$  &$EF^{1/2}_k$, $k=3,4$&0&&ah\\ \hline
 &$EF^{1/2}_3F^{1/2}_4$&3&1&h\\ \hline
&$EF^{1/2}_kF^{1/2}_5$&3&1&h\\ \hline
&$EF^{1/2}_3F^{1/2}_4F^{1/2}_5$&1&5&h\\ \hline \hline
$(J=1,10/\bar{10})$ &$EF^{1/2}_kF^{1/2}_l$, $k\neq l=3,4,5$&1&2&$\frac{3}{p}$
\\ \hline
$(J=0,10/\bar{10})$ &$F^{1/2}_3F^{1/2}_4F^{1/2}_5$&1&2&$\frac{3}{p}$
\\ \hline
$(J=0,10/\bar{10})$ &$G^{3/2}_5$&1&0&0\\ \hline\hline
 $(J=1,27)$ &$EF^{1/2}_k$, $k=3,4,5$&1&1&$h$
\\ \hline
$(J=0,27)$ &$F^{1/2}_kF^{1/2}_l$, $k\neq l=3,4,5$&1&1&$h$
\\ \hline\hline
 $(J=1,27)$ &$E$&1&0&$0$\\ \hline
$(J=0,27)$ &$F^{1/2}_k$, $k=3,4,5$&1&0&$0$\\ \hline\hline
\end{tabular}

\vl

Table 6.2. \label{Exodege} Degeneracies of $M^2=0,3/2$ and (in some cases)
$M^2=3$ states for    exotic bosons.
  Exotic bosons are characterized by operator $O$ creating the  (in
general  tachyonic)  ground state for the boson.  The masses are given for
$T=1$.  'ah' (' absolutely heavy') means that state has Planck
 mass  independentely of the value of $p$  and $T$ and  'h' (¬heavy ')
 means that state has Planck mass for  primes, which are not too large
and for $T=1$: for $T=1/2$ state has mass of order $1/p$. No massless
exotic  noncolored bosons are predicted.   There are however some massless
colored states. Note that $F^{1/2}_k$ contains isospin index and  the case
 of intermediate gauge bosons suggests certain
 constraints  on isospin indices.

\subsection{   Photon}

Photon and $Z^0$ differ in one respect only: the charge matrix for photon
 acts in the sectors $k=3,5$ and for $Z^0$ in sectors $k=3,4,5$. Photon
and $Z^0$ are superpositions of states $k$ is 'active' that is N-S vacuum
 with $h_k=1/2 $ is  excited. The 3 sectors decouple from each other and
apart from the presence of nonzero norm states created by $L^2_5$ and
$G^{k}, k>3/2$ from $h_5=0$ vacuum the sectors  are essentially identical.

\vm

The ground state for  photon is given  by the following
expression

\begin{eqnarray}
\vert vac \rangle_{\gamma}&=& E (aF^{1/2,3}_3+ bF^{1/2}_5)
\vert vac \rangle\nonumber\\
\end{eqnarray}

\noindent where the values of the coefficients $a$ and $b$ are such that
 the coupling $Q_{em}=I_3+Q_K/2$  to fermions results.  State satisfies
gauge conditions.  For definitess it will be assumed that polarization
operator acts in sector $i=2$.   Note that  $T0_3$ and $ T^0_5$  and also
their super counterparts appearing in the state measure vectorial isospin
and K\"ahler charge.

\vm

{\bf 1. Massless states}

\vm

Massless photon state is obtained by applying operators $O^{3/2}$  to the
ground state. Polarization operator $P$  is not allowed in the
construction.    State contains two terms of same form corresponding to
$k=3,5$ charge operators and these terms do not couple to each other in
gauge conditions. Therefore one can consider only the second term.  The
list of these operators  for say $k=3$ case is following:\\ a) Single
particle operators $O^{3/2}_i$,$ i=2,3$:  $2+2=4$ altogether.\\ b)  Two
particle operators
 $O^{1}_i O^{1/2}_j$, $i\neq j=2,3$ ($2$).\\ c)  $G^{3/2}_5$ gives one
additional operator in $k=3$ case. Gauge conditions are identically
satisfied for this operator.\\ There are $7$  ($6$) $O^{3/2}$ operators
altogether for $k=3$ ($k=5$).

\vm

Gauge conditions for $G^{1/2}$  in case of $k=3$ correspond to operators
$O^1$
 given by  $O^1_i$, $i=2,3$ ($2$)  and $O^{1/2}_2O^{1/2}_3$:  $3$
altogether. For $L^1$ gauge conditions correspond to the  $2$ operators
 $O^{1/2}$ given by $O^{1/2}_i$ , $i=1,2$.  Gauge conditions for $L^2$ are
identically satisfied.  This gives $5$ conditions altogether  so that
$k=3$ gives $D(0,3)=2$   states and $k=5$ $D(0,5)=1$ state.  Ground state
degeneracy for photon is $D(0,\gamma)=2+1=3$.

\vm

If the  inner product is modified then  $2$ additional operators
$O^{3/2}_i$,
 $i=1,4$ are allowed in $N=1$ sector for $k=3$. For $k=5$ therere are $3$
operators. Gauge condition associated with these operators are trivially
satisfied so that $5$ states are obtained.    Ground state degeneracy
becomes $D(0,\gamma)=3+5=8$.

\vm

{\bf 2.  $M^2=3/2$ states}

\vm

Massive states at level $\Delta=1$ are created by operators
 $O^{5/2}$ from ground states.  Again conditions are independent for $k=3$
and $k=5$ contributions to the charge operator. For $k=3$ the list  of
excitations not involving color is following:\\ a)  Single particle
operators $O^{5/2}_i$, $i=2,3$ (8).  \\  b)  Two-particle operators
$O^{2}_iO^{1/2}_j$ ($3+3=6$),   $O^{3/2}_i O^{1}_j$ ($2+2=4$), $i\neq
j=2,3$: $10$ altogether.  \\ c) The $6$ operators $G^{5/2}_5$,
$G^{1/2}_iL^2_5$ and   $L^1_iG^{3/2}_5$ and
 $G^{3/2}_5G^{1/2}_2G^{1/2}_3$, $i=2,3$ acting in $u(1)$ degrees of
freedom  are present only for $k=3$.\\ d)  The operators $O^{5/2}_c$(1,
see for table \ref{Nsginvdege} in appendix F) and
 $O^{2}_cG^{1/2}_i$ (2), $i=1,2$ acting in color degrees of freedom.
 There are $3$ operators altogether. \\ The total number of operators is $
8+10+6+3=27$ for $k=3$ and $21$ for $k=5$.

\vm

a) Gauge conditions for $G^{1/2}$ ($k=3$)  correspond to the   operators
$O^{2}$   given by \\ i) $ O^2_i$,$i=2,3$ ($3+3=6$)\\ ii)
$O^{3/2}_iO^{1/2}_j$, $O^1_iO^1_j$($5$), $i\neq j=2,3$.\\ ii) $L^2_5,
G^{3/2}_5G^{1/2}_i$, $i=2,3$ present only for $k=3$ (3).    \\ iii)
$O^2_c=F^{A3/2}F^{A1/2}$.  (1) \\ There are $6+5+3+1= 15$ conditions for
$k=3$ and $12$ conditions for $k=5$. \\ b) For $L^1$ the gauge conditions
correspond to the operators $O^{3/2}$ and
 their number is $7$ for $k=3$ and $6$ for $k=5$.\\  c) For  $G^{3/2}$
gauge conditions correspond to the $2$ operators $O^{1}_i$,
 $ i=2,3$.  For $L^2$ gauge conditions correspond to unit matrix. \\ The
total number of gauge conditions is $15+7+2+1=25$ for $k=3$ and
$12+6+2+1=21$ for $k=5$. The contribution to degeneracy is $D(1,3)=27-25=2$
 for $k=3$ and $D(1,5)=21-21=0$ for $k=5$ so that the degeneracy is
$D(1,\gamma)=2$.  Since $D(0,\gamma)=3$ photon is essentially massless
provided $k(B)$ is multiple $3/2$.

 \vm

In $N=1$ there are  operators $G^{5/2}_i$ , $L^2_iO^{1/2}_j$,
$G^{3/2}_iO^1_j$,$G^{3/2}_iO^{1/2}_2O^{1/2}_3$   $j=2,3$,  $i=1,4$  for
$k=3$:
 their total number is $11$.  Gauge conditions for $G^{1/2}$ correspond
to  $6$ operators $L^{2}_i$ , , $G^{3/2}_iO^{1/2}_j$,   $j=2,3$,  $i=1,4$
for $k=3$. For  $L^1$ and $L^2$ gauge conditions correspond to the
operators $ G^{3/2}_i$ (2) and $O^{1/2}_j$,$j=2,3$ (2). The total number
of gauge conditions is $10$ so that one obtains one  state for $k=3$.
For  $k=5$ the number of states is $2$. This
 implies $D(3/2)/D(0)=(2+3)/8\neq 1$.

\vm

The conclusion is that photon possesses negligibly small mass  for
$T(ew)=1/2$, $k\ge 2$ provided the value of the parameter $k(B)$ is
$k(B)=3/4$.  At level $M_{127}$ this mass would be  of order
$m_{e}/\sqrt{M_{127}} \simeq 10^{-13} \ eV $.  The results apply almost as
such to the case of $Z$ boson.

\subsection{$Z^0$ boson}

The ground state for $Z^0$ is  given by the following
expression

\begin{eqnarray}
\vert vac \rangle_{Z}&=& E(aF^{1/2,3}_3+ bF^{1/2}_4+cF^{1/2}_5)
\vert vac \rangle\nonumber\\
\end{eqnarray}

\noindent where the coefficients are determined from the requirement  that
$Z^0$ couplings to are given by $Q_Z= I^3_L+ sin^2(\theta_W) Q_{em}$.  What
differentiates between $Z^0$ and $\gamma$ is the fact that the  coupling
is not purely vectorial for $Z^0$ and three active sectors $k=3,4,5$
become possible.  The degeracies for various states can be obtained using
the results for photon. The total degeneracy for level $n$ is given by

\begin{eqnarray}
D(n)&=& \sum_{k=3,4,5}D(n,k)= 2D(n,3)+D(n,5)
\end{eqnarray}

\noindent The degeneracies $D(n,3)=D(n,4)$ and $ D(n,5)$ have been
 already calculated:

\begin{eqnarray}
D(0,3)&=&2, \ \ D(0,5)=1\nonumber\\
D(1,3)&=&2, \  \  D(1,5)=0\nonumber\\
\end{eqnarray}
\noindent so that one has for the degeneracies of various states

\begin{eqnarray}
D(0,Z)&=&5 \nonumber\\
D(1,Z)&=&4
\end{eqnarray}

\noindent The ratio $s(Z)=k(B)D(1,Z)/D(0,Z)=n6/5$ ($k(B)=n3/2$ from the
 masslessness of photon)  implies that $Z^0$ must correspond to temperature
$T(Z)=1/2$.  $n=1 $ turns out to be the only physically interesting values
of $n$.  A small calculation shows that the real counterpart of
$p^2/5$($n=4$) is $4/5$, which corresponds to $s(Z)=0$ and $X= 12$ in
small quantum number approximation.

\subsection{  $W$ boson}

The ground state for  $W$  boson is given by the following
expressions

\begin{eqnarray}
\vert vac \rangle_{W}&=& E \sum_{i,j=1,2}
 q_{ij}F^{1/2,i}_3F^{1/2,j}_4\vert vac \rangle\nonumber\\
\end{eqnarray}

\noindent The coefficients $q_{ij}$ are determined by the charge matrix of
$W$.
 The difference between $Z^0$ and $W$ is that the sectors $i=3,4$ are both
active simultaneously.

\vm

  Massless states are created by applying operators $O^{1}$ to  the $W$
ground state. The list of $O^1$ operators is following.\\ a) Single
particle operators $O^1_i$, $i=2,3,4$: $3$ altogether. \\ b) Two particle
operators $O^{1/2}_iO^{1/2}_j$, $i\neq j=2,3,4$: $3$
 altogether.\\ The total number of operators is $3+3=6$.  $G^{1/2}$ gives
$3$
 gauge conditions (operators $O^{1/2}_i$) . $L^1$ gives one gauge
condition,
 which is  however not an independent one due to the antisymmetry of the
coefficient matrix associated with $O^{1/2}_iO^{1/2}_j$. Therefore  ground
state degeneracy is $D(0,W)=6-3=3$.

\vm

$M^2=3/2$ states are created by applying operators $O^{2}$ to the $W$
ground  state. The list of $O^2$ operators is following.\\ a) Single
particle operators $O^2_i$, $i=2,3,4$: $9$ altogether. \\ b) Two particle
operators $O^{3/2}_iO^{1/2}_j$ (12), $O^{1}_iO^{1}_j$ (3),
  $i\neq j=2,3,4$: $15$ altogether.\\ c) 3-particle operators
$O^{1}_iO^{1/2}_jO^{1/2}_k$, $i\neq j\neq k=2,3,4$:  $3$ altogether. \\ d)
Operators $ L^2_5, G^{3/2}_5O^{1/2}_i$,$i=2,3,4$ acting in $u(1)$ degrees
of freedom: $4$ altogether.\\ e) Operator $O^2_c$ acting in color degrees
of freedom. \\ The total number of operators is $9+15+3+4+1=32$.

\vm

Consider next gauge conditions.\\ a) $G^{1/2}$ gauge conditions correspond
to\\ i) single particle operators $O^{3/2}_i$, $i=2,3,4$: $6$ altogether.
\\ ii) two-particle operators $O^{1}_iO^{1/2}_j$ ,  $i\neq j=2,3,4$: $6$
altogether.\\ iii) 3-particle operator $O^{1/2}_1O^{1/2}_2O^{1/2}_3$. \\
iv) operator $ G^{3/2}_5$.\\ There are $6+6+1+1=14$ operators altogether.
\\ b)$L^1$ gauge conditions correspond \\ i) single particle operators
$O^{1}_i$, $i=2,3,4$: $3$ altogether. \\ ii) two particle operators
$O^{1/2}_iO^{1/2}_j$ ,  $i\neq j=2,3,4$: $3$  altogether.\\ $3+3=6$
conditions altogether. \\ c) $L^2$ gives one condition.\\ The total number
of gauge conditions is $14+6+1= 21$ and the
 degeneracy is $D(1,W)=32-21=11$.   $W$ mass is is of order Planck mass
unless one assumes $T(W)=1/2$.   In this case  second order contribution
is $n 11/2$ and  one has $s(W)=0$ and $X(W)= 8$ for $n=1$.

\vm

The value of Weinberg angle serves as test for the physicality of the
scenario.  For $n=1$ ($k(B)=3/2$) one obtains

\begin{eqnarray} sin^2(\theta_W)&=&1-\frac{ M_W^2}{M_Z^2}=\frac{3}{8}
\end{eqnarray}

\noindent Encouragingly, the value is typical value of the parameter in the
 symmetry limit in GUTs. Electron gauge boson mass ratio comes out
correctly if one assumes $k(F)=2$.

\subsection{  Gluon}

There are no color octet operators $O^{kA}_c$  for $k=1,3/2,....,4$
satisfying the gauge conditions associated with $L^1$, $G^{1/2}$ and
$L^2$  as  the study of the table
 (see \ref{Nsginvdege} in appendix F) shows.  This means that   a natural
identification for the gluon ground state is as the state created
 by the operator

\begin{eqnarray}
G&=&  EF^{A1/2}\nonumber\\
E&=& \epsilon_k \gamma^k_{1/2}
\end{eqnarray}

\noindent  The interesting physical states are created by operating with
 $so.. \times u(1)$ operators on this state.  Massless states are created
by operating with the operators $O^{3/2}_i $, $i=2,5$  to this state. Gauge
conditions for $G^{1/2}$ and $L^{1}$ leave only $G^{3/2}_5$ so that the
ground state degeneracy is $D(0,G)=1$.

\vm

For $M^2=1$ state the $7$  operators $O^{5/2}$ are given by $O^{5/2}_2$,$
G^{5/2}$,   $L^2_5G^{1/2}_2$ and $G^{3/2}_5L^1_2$. The number of gauge
conditions is $9$ corresponding to operators $ O^2_2$,$ L^2_5$,
 $ G^{3/2}_5F^{1/2}_2$ (5) associated with $G^{1/2}$, to the operators
$O^{3/2}_2, G^{3/2}_5$ (3) corresponding to  $L^1$ and  operator $O^1_2$
corresponding to $L^2$. There are no solutions to  gauge conditions so
that gluon is exactly massless in order $O(p)$.  Since $D(0,G)=1$ the next
order gives negligible contribution to mass.

\section{Appendix E: Exotic states }

 There are many candidates for exotic particles and one must show that
these
 states  \\ a) do not allow massless ground state satisfying gauge
conditions or \\ b) (assuming that $T=1$) possess Planck mass
($D(3/2)/2D(0)$ is not an integer)
 or\\ c) have nonvanishing $D(1)$ or at least  $X \ mod \ 2D(0)\neq 0$,
 $X= 3(2D(3)-D(3/2)^2/D(0))$ so that they become massive ($k(B)=3/2$ is
assumed in the formula).

\vm

Exotic states can be classified according to the properties of the ground
 state associated with the particle. \\ a) Exotic scalars with ground
state,  which can have nonvanishing electroweak
 quantum numbers. \\ b) Scalar gluons. \\ c) Exotic spin 1 electroweak
bosons \\ d) Exotic spin 1 gluons.

\subsection{Exotic scalars}

It turns out that there are many exotic scalars allowing massless
excitation but that for $T=1$ no exotic scalar remains massless in
Higgs mechanism. For $T=1/2$ there are two massless scalars.
 These states are created by the following operators.

\vm

a) $ O=1$: Planck mass\\
  Massless states are created by operators $G^{5/2}$ and $O^{5/2}_c(1)$.
Gauge  conditions allow no massless excitations.

\vm

b) $O=F^{1/2}_5$: Planck mass\\ There are no operators creating massless
states and state possesses Planck mass.

\vm

c) $O=F^{1/2}_k$, $k=3,4$: Planck mass for $T=1$. \\
  The operators $O^2$
creating massless states are \\ i) single particle $O^2_k(3)$ and
operators
$L^2_5$(1),
 $G^{3/2}_5G^{1/2}_k$(1)\\ ii) the operator  $O^2_c$(1) creatig color
excitation. \\ There are $6$ operators altogether.

\vm

c1) Gauge conditions of $G^{1/2}$ correspond to  single particle operators
 $O^{3/2}_k$(2)
 and operator $G^{3/2}_5$: $3$ altogether.\\
 c2) Gauge conditions $L^1$ correspond to   operator $O^1_k$(1)\\ c) Gauge
conditions for $L^2$ corresponds to unit operator.\\ Total number of gauge
conditions is $5$ so that $D(0)=1$ results. The states can be regarded as
scalar counterparts of axial and vectorial gauge bosons with $I_3=0$.
$D(0)=1$ implies that these have Planck mass    for $T=1$ and $k(B)=3/2$
if $D(3/2)$ is odd. It turns out that $D(3/2)=1$!

\vm

 Because of its  central importance the result $D(3/2)=1$  will be derived by
considering opertors $O^3$  creating $\Delta=1$  states. Consider first
$O^3$ operators creating $M^2=3/2$ states.  There are\\ a) single particle
operators $O^3_k$(5) and operators $L^3_5$, $G^{5/2}_5G^{1/2}_k$(1),
$L^{2}_5O^{1}_k$(1) $G^{3/2}_5O^{3/2}_k$(2): $10$ altogether.\\ b) The
operators $O^3_c$(2) $O^{5/2}_cO^{1/2}_k$(1),  $O^2_cO^1_k$(1)  creating
color excitations. There are $4$ operators altogether.

\vm

Gauge conditions allow no operators of type a):\\  a) Gauge conditions of
$G^{1/2}$ correspond to   single particle operators
 $O^{5/2}_k$(4) and operators $G^{5/2}_5$,
 $L^{2}_5G^{1/2}_k$(1), $G^{3/2}_5O^{1}_k$(1): $7$ altogether.\\ b) Gauge
conditions $L^1$ correspond to   operators $O^2$: their total
 number is  $5$\\ c) Gauge conditions for $L^2$ correspond to operator
$L^1_k$. \\ Total number of gauge conditions is $7+5+1=13>10$ so that no
solutions to gauge
 conditions are possible. \\ For operators of type b)  gauge conditions
for $G^{1/2}$ correspond to operators
  $O^{5/2}_c$(1) and
 $O^{2}_cO^{1/2}_k$(1). Gauge conditions for $L^1$ corresponds to the
operators
 $O^2_c$. $L^2$ gives no conditions. The number of gauge conditions is
 therefore $3$ and just one operator satisfying gauge conditions remains
and one has $D(3/2)=1$.

\vm

c) $O= F^{1/2}_kF^{1/2}_l$, $k,l= 3,4,5$.\\
 $(k,l)=(3,4)$: Planck mass for $T=1$.\\
$(k,l)=(3,5),(4,5)$: $\frac{3}{p}$ for $T=1$.\\
For $k=3,l=4$ the ground state
degeneracy is $D(0)=2$ and for  $(k,l)=(3,5),(4,5)$ the ground state
degeneracy is $D(0)=1$. These states can be regarded as scalar partners of
$W$ and as a   $u(1)$ boson coupling to the product of axial/vectorial
isospin and K\"ahler charge. The value of $D(3/2)$  is $3$ for
 $(k,l)=(3,4)$
so that this state has Planck mass.   $D(3/2)=2$ for $l=5$ boson implies
that this state has mass doesn't produce troubles
 if it condenses on level with  small $p$. For $T=1/2$ the state is
essentially massless.

\vm

d) $O= F_3^{1/2}F_4^{1/2}F_5^{1/2}$: Planck mass for $T=1$.\\
 Massless ground
state degeneracy is $D(0)=2$.  State couples to the product of
$I^{\pm}$ and
K\"ahler charge. $D(3/2)=10$ implies that state has Planck mass for
 $T=1$ and mass $M^2=\frac{1}{2p}$ for $T=1/2$.

\subsection{Exotic scalar gluons}

There is only one exotic scalar gluon.\\ a) $O= F^{A1/2}$,
$F_k^{1/2}F^{A1/2}$ and  $F_{3}^{1/2}F_{4}^{1/2}F^{A1/2}$ allow no
 massless states and possess therefore Planck mass.

\vm

\noindent b) $O= F^{1/2}_3 F^{1/2}_4F^{1/2}_5F^{1/2A}$: $M^2=
\frac{2}{p}$ for $T=1$.\\ There are $D(0)=3$ massless states.
 State couples to $I^{\pm} Q_K T^A$.  $D(3/2)=4$ implies that state is
 not massless for $T=1$ although it is 'light': for $T=1/2$ state would
be massless.

\subsection{Exotic spin one color singlet bosons}

The operator creating these states is of form $B= EO$.
\\

\vm

 $O=1$: Planck mass for $T=1$.\\
This boson couples to fermion number.
 Massless ground state has $D(0)=1$     $D(3/2)=1$
follows using the same  derivation as applied in case of scalar bosons
$(O=F^{1/2}_k$) so that state has Planck mass  for $k(B)=3/2$ and $T=1$.

\subsection{Exotic spin one gluons}

Again states are of form $B=EO$.\\
a) $O= F^{1/2}_kF^{A1/2}$: Planck mass.\\
 Gauge conditions allow no massless ground states.

\vm

\noindent b) $O= F^{1/2}_kF^{1/2}_lF^{A1/2}$, $k,l=3,4,5$: Planck mass for
$T=1$.\\
 Massless ground state degeneracy is  $D(0)=3$.  These states include
color octet counterpart of $W$ boson. These states couple to $I^{\pm} T^A$
or to $I^3_{A/V}Q_K$. $D(3/2)=1$ makes these states very massive for
 $k(B)=3/2$
and $T=1$.

\vm

\noindent c) $O= F^{1/2}_3F^{1/2}_4F^{1/2}_5F^{A1/2}$: Planck mass for
 $T=1$.\\
 Massless ground state degeneracy is $D(0)=1$.
  $D(3/2)=5$ implies that this state has Planck mass for $k(B)=3/2$ and
$T=1$.

\subsection{Higher color representations}

The table  of appendix F for  gauge invariant N-S type  color operators
with various conformal weights $n$  shows the existence of $n=1$ decuplets
 $10$  and $\bar{10}$,  $n=3/2$  27-plet and doubly degenerate  $n=2$
27-plet.   This  implies the existence of  exotic colored bosons.
Also p-adic temperature $T=1/2$  is possible  for these
states since the existence of very light colored states does not imply new
long range forces.

\vm

a) Decuplet operators  creating massless states have the general form
$O = O^{3/2} O^{10,1}_c$, where $O^{3/2}$ acts in $so...\times u(1)$ degrees
of freedom. There are the following possibilities: \\ i)  For
$O^{3/2}=EF^{1/2}_kF^{1/2}_l$ , $  k\neq l =3,4,5$  and
$O^{3/2}=F^{1/2}_3F^{1/2}_4F^{1/2}_5$   gauge conditions are identically
satisfied and one has $D(0)=1$.  $M^2=3/2$ excitations are created by the
3  operators $L^1_i,i=2,k,l$ (3,4,5)  and $L^1$ gauge condition implies
that degeneracy of $M^2=3/2$ states is $D(3/2)=2$  so that   for $T=1$
state has mass $M^2= \frac{3}{p}$.  For $T=1/2$ the state is essentially
massless. \\ ii) $O^{3/2}=G^{3/2}_5$.  Gauge conditions are identically
satisfied. There are neither  $M^2=3/2$ nor  $M^2= 3$ excitations
satisfying gauge conditions so that these states are exactly massless. \\
iv) There are also operators of form $O^{3/2}= O^{n_1}O^{n_2}$,
$n_1+n_2=3/2$,   where $O^{n_1}$ is Super Virasoro generator and $O^{n_2}$
is constructed from the operators $ O^{1/2}_i$ but these operators give no
states satisfying gauge conditions.

\vm

b)  The  operators creating massless states in case of $O^{27,3/2}$ have
 the general form $O=O^1O^{27,3/2}_c$.  Only  $O^1= EF_k^{1/2}$ , $k=3,4,5$
and $O^1= F_k^{1/2}F_l^{1/2}$ give  states satisfying  gauge conditions and
degeneracies are $D(0)=1$  and $D(3/2)= 1$.  These states have Planck mass
for $T=1$ and for $T=1/2$ the mass is $M^2=\frac{1}{2p}$.

\vm

c)  The  operators creating massless states in case of $O^{27,2}$ have
 the general form $O=O^{1/2}O^{27,2}_c$  with $O^{1/2}= E$ or
 $O^{1/2}= F_k^{1/2}$, $k=3,4,5$  and there exists double-fold
degeneracy.    The  degeneracies are $D(0)=1$  and $D(3/2)=
0$ and the states are massless  for both $T=1$ and  $ T=\frac{1}{2}$.

\section{ Appendix F: Construction of positive norm states in color
 degrees of freedom}

The construction of positive norm states for various values of conformal
weight is essential ingredient in the calculation of degeneracies for
various values of mass squared operator in order to estimate thermal mass
expectation value: \\ i) If one has obtained the multiplicities of various
representations with weight $n$ then it is easy to calculate the
multiplicities for gauge invariant states. If gauge conditions associated
with $G^{1/2}$, $L^1$  and $L^2$  in N-S representation induce surjective
maps to the levels $n-1/2$,$n-1$ and $n-2$ then the multiplicity of gauge
invariant  representation is given by $m= m(n)-m(n-1/2)-m(n-1)-m(n-2)$.
\\ ii) For states involving tensor product of several Super Virasoro
representations (say, the representation of baryon as 3-quark state)  it
is straightforward task to form tensor product in color degrees of freedom
if multiplicities of nonzero norm states are known and apply then gauge
conditions to the total Super Virasoro.

\vm

The construction of nonzero norm states relies on the following
observations. \\ a) The states at each level $n$  (conformal weight)  of
color Kac Moody
 algebra can be classified into irreducible representations of color
group. The states are created by polynomials  $O(F)$, where $F$ is short
hand notation  for the super generators $F^{Ak}$ of color  Kac Moody
algebra.  Bosonic generators are not used since they are expressible in
terms of $F^{Ak}$ and their use would lead to double counting problems
since $T^{An}$ is expressible as bilinear of $F$.\\ b) Zero norm states
must be eliminated.  They are created by product  operators of form

\begin{eqnarray}
O&=& O_1O_2\nonumber\\
O_1&=& O(F)\nonumber\\
O_2&=& O_2(L,G)\nonumber\\
\end{eqnarray}

\noindent $O_2$ is operator formed from Super Virasoro generators and
creates  zero norm state ($c=h=0$).  $O_1$ is polynomial of fermionic
generators $F^{Ak}$ in color Kac Moody algebra. \\ c) The  operators $O_2$
are constructed from
 Super Virasoro generators,
 which do not annihilate vacuum state.  For N-S algebra there are
generators, which annihilate vacuum automatically and must be excluded
from construction.  The generators $G^{1/2}$ and $L^1$ are proportional to
$T^{A0}$ and annihilate therefore vacuum. The generator $L^{2}$ reduces
essentially  to $T^{1A}T^{1A}$, when acting on vacuum. The representation
$T^{1A}\propto f^{ABC}F^{1/2B}F^{1/2C}$ together with Jacobi identities
demonstrates that the action of  $T^{1A}T^{1A}$ to vacuum  actually
vanishes.  Since $L^1$ and $L^2$ generate  Virasoro algebra all generators
$L^n$ annihilate vacuum so that only the operators  constructed from
$G^k,k>1/2$ remain to be considered.   For Ramond representations the
entire Super Virasoro algebra must be taken into consideration since
$T^{A0}$ do not annihilate ground state triplet.  \\ d) The operators
$O_1$ creating nonzero norm sates can be classified into irreducible
representations
 of color group. The basic building blocks are the representations defined
by  $N$:th order  monomials of generators $F^{Ak}$ with $k$ fixed.  These
representations are completely antisymmetrized tensor products of
 $N=0,1,....,8$ octets   and representation content is same for all values
of $k$.  The representation content can be coded into multiplicity vector
$m(N;k)$,  $k=1,8,10,....$,\\  e) Once the representation contents for
antisymmetrized tensor
 products are known  in terms of multiplicity vectors,  the representation
 contents for tensor products of $N_1,k_1$ and $N_2,k_2$ can be determined
by standard tensor product construction since anticommutativity  does not
produce no effects for $k_1\neq k_2$.  One can express the multiplicity
vector for the tensor product $(N_1,k_1)\otimes (N_2,k_2)$ in terms of the
multiplicity vector $D(k_1,k_2,k_3)$ for the tensor product of irreducible
representations  $k_1,k_2 =1,8,10,...$.

\begin{eqnarray}
m((N_1,k_1)\otimes (N_2,k_2;k)&=& m(N_1;k_1)D(k_1,k_2,k_3)
m(N_2;k_2)
\end{eqnarray}

\noindent f)  It is useful to caculate total multiplicity vector $m(n;k)$
for each
 conformal weight $n$ by considering all possible states having this
conformal weight.   The multiplicity vector is just the sum of
multiplicity vectors of various tensor products satisfying  $\sum N_ik_i=
N$:

\begin{eqnarray}
m(n;k)&=&\sum_{S=N} m((N_1,k_1)\otimes ....\otimes (N_r,k_r; k))\nonumber\\
S&\equiv& \sum N_ik_i
\end{eqnarray}

\noindent  The multiplicity vectors $m(n;k)$  are basic objects in the
 systematic construction of tensor products of several  Super Virasoro
algebras (say, in construction of many quark states).

\subsection{Multiplicity vectors for antisymmetric tensor products}

Consider first the construction of $N$-fold  antisymmetric tensor products
of  octets $F^{Ak}$, $k$ fixed. The tensor products are obviously
analogous to the antisymmetric tensors of $8$-dimensional space.
   The completely antisymmetric 8-dimensional permutation symbol
$\epsilon_{A_1,....,A_8}$ transforms as color singlet and induces duality
operation in the set of antisymmetric representations: the antisymmetric
representations  $N$ are mapped to representations $8-N$.  This imlies
that the representation contents are same for $N=0$ and  $8$, $N=1$ and
$7$, $N=2$ and $N=6$, $N=3$ and $N=5$ respectively. $N=4$ is self dual.
It is relatively easy to determine the representation content of the
lowest completely antisymmetric representations and the results can be
summarized conveniently as multiplicity vectors defined as

\begin{eqnarray}
\bar{m}&\equiv& (m(1),m(8),m(10),m(\bar{10}),m(27), m(28),m(\bar{28}),
 m(64), m(81),m(\bar{81}), m(125),...)\nonumber\\
\
\end{eqnarray}

\noindent The multiplicity vectors are given by the following formulas

\begin{eqnarray}
\bar{m}(F)= \bar{m}(F^7)&=& (0,1)\nonumber\\
\bar{m}(F^2)=\bar{m}(F^6)&=& (0,1,1,1)\nonumber\\
\bar{m}(F^3)=\bar{m}(F^5)&=& (1,1,1,1,1)\nonumber\\
\bar{m}(F^4)&=& (0,2,0,0,2))
\end{eqnarray}

\noindent where $F^N$ denotes N:th tensor power of $F^{Ak}$.

\subsection{Multiplicity vectors for various conformal weights for color
 Super Virasoro algebra}

The next task is to calculate multiplicity vectors for various conformal
 weights. The task is straightforward application of Young Tableaux . The
representation contents  for various conformal weights for N-S algebra are
given by

\begin{eqnarray}
n&=&0: 1\nonumber\\
n&=&1/2: 1/2 \nonumber\\
n&=&1: (1/2)^2)\nonumber\\
n&=&3/2: 3/2\oplus (1/2)^3\nonumber\\
n&=&2: (3/2)\otimes (1/2)\oplus (1/2)^4\nonumber\\
n&=&5/2: 5/2\oplus (3/2)\otimes (1/2)^2\oplus (1/2)^5\nonumber\\
n&=&3: (5/2)\otimes (1/2) \oplus (3/2)^2\oplus (3/2)\otimes (1/2)^3
 \oplus (1/2)^6\nonumber\\
n&=&7/2:  7/2\oplus (5/2)\otimes (1/2)^2\oplus (3/2)^2 \otimes (1/2)
\oplus 3/2\otimes (1/2)^4\oplus (1/2)^7\nonumber\\
 n&=&4: (7/2)\otimes (1/2) \oplus (5/2)\otimes (3/2)\oplus (5/2)\otimes
(1/2)^3  \oplus    (3/2)^2\otimes (1/2)^2...\nonumber\\
&\oplus& (3/2)\otimes
(1/2)^5  \oplus (1/2)^8\nonumber\\ n&=&9/2:  9/2\oplus (7/2)\otimes
(1/2)^2...\nonumber\\ &\oplus&
(5/2)\otimes (3/2)\otimes (1/2) \oplus 5/2\otimes
 (1/2)^4\oplus (3/2)^3....\nonumber\\
&\oplus& (3/2)^2\otimes (1/2)^3\oplus (3/2)
\otimes (1/2)^6\nonumber\\
\
 \end{eqnarray}

\noindent Multiplicity vectors  obtained as sums of multiplicity vectors
 associated with summands in the direct sum composition and are given by
the following table

\vl

\begin{tabular}{||c|c|c|c|c|c|c|c|c|c|c|c|c||}
 \hline\hline
n&1&8&10&$\bar{10}$&  27&28&$\bar{28}$&35&$\bar{35}$& 64&81&$\bar{81}$\\
\cline{1-13}\hline
 0    &1&    &  &       & &&&&&  &&\\ \hline
1/2&  &1     &  &       & &&&&&  &&\\ \hline
1      &     &1 &1&1    &  &&&&&   &&\\ \hline
3/2&1&2  &1&1   &1&&&&& &&\\ \hline
2    &1&4  &1&1   &3 &&&&& &&\\ \hline
5/2&2&6  &3&3   &4 &&&&& &&\\ \hline
3    &2&10&6&6&6  &&&2&2&1 &&\\ \hline
7/2&4&16&8&8&12&&&4&4&2&&\\ \hline
4    &8&24&12&12&21&1&1&7&7&4&&\\ \hline
9/2&10&36&21&21&32&1&1&12&12&8&1&1\\ \hline \hline
\end{tabular}

\vl

Table 8.1. \label{Nscolordege} Multiplicity vectors for various conformal
 weights for N-S type Super
Virasoro algebra.

\vm

Similar arguments can be used to deduce the multiplicity vectors in case of
 Ramond type Super Virasoro algebra.

\begin{tabular}{||c|c|c|c|c|c|c|c|c|c|c|c|c|c|c|c||}
 \hline\hline
n&1&8&10&$\bar{10}$&  27&28&$\bar{28}$&35&$\bar{35}$&
 64&80&$\bar{80}$&81&$\bar{81}$&125\\
\cline{1-16}\hline
 0    &1&    &  &       & &&&&&  &&&&&\\ \hline
1&  &1     &  &       & &&&&&  &&&&&\\ \hline
2      &     &1 &1&1    &  &&&&&   &&&&&\\ \hline
3&2  &4  &2&2   &2&&&&& &&&&&\\ \hline
4 &2 &10  &4   &4   &6    &  &  &1  &1  &  &&&&&\\ \hline
5&6  &20  &10&10  &14  & &   &4  &4  &1 &&&&&\\ \hline
6&12&40  &22&22  &32  &1&1&10&10&6 &&&&&\\ \hline
7&17&68  &36&36  &55  &1&1&20&20&11& & &1&1&\\ \hline
8&33&124&70&70&113  &5&5&44&44&29& & &5&5&1\\ \hline
9&70&276&170&170&276&16&16&122&122&94&1&1&22&22&6\\ \hline \hline
\end{tabular}

\vl

Table 8.2.  \label{Racolordege} Multiplicity vectors for various conformal
 weights for Ramond type Super Virasoro algebra.

\subsection{Elimination of zero norm state from N-S and Ramond  algebra}

Multiplicity vectors for the zero norm representations in  N-S algebra  can
 be constructed easily. There are two important point to notice.\\  i) The
operators  $G^{1/2}$, $L^n$ annihilate vacuum and cannot be used in
 the construction of module of zero norm states.\\ ii)  The polynomial
$O_1$ of color Kac Moody  super generators multiplying
 Virasoro operator $O_2$ must have  nonvanishing norm in order to avoid
double counting. \\
 ii) The operators are most conveniently constructed by starting from
$n=0$  level and proceeding iteratively counting simultaneously the
multiplicity vectors for nonzero norm states.\\ The list is of  Super
Virasoro operators generating zero norm states  and elimination  procedure
is described in following:

\begin{eqnarray}
n&=&3/2: G^{3/2}  \nonumber\\
\bar{m}(3/2)&\rightarrow& \bar{m}(3/2)-\bar{m}(0)\nonumber\\
n&=&2:G^{3/2} \nonumber\\
\bar{m}(2)&\rightarrow& \bar{m}(2)-\bar{m}(1/2)\nonumber\\
n&=&5/2: G^{5/2}, G^{3/2} \nonumber\\
\bar{m}(5/2)&\rightarrow&\bar{m}(5/2)-\bar{m}(0)- \bar{m}(1)  \nonumber\\
n&=&3: G^{5/2}, G^{3/2} \nonumber\\
\bar{m}(3)&\rightarrow&\bar{m}(3)-\bar{m}(1/2)- \bar{m}(3/2)
\nonumber\\ n&=&7/2: G^{7/2},G^{5/2}, G^{3/2} \nonumber\\
\bar{m}(7/2)&\rightarrow&\bar{m}(7/2)-\bar{m}(0)- \bar{m}(1)-
\bar{m}(2)           \nonumber\\ n&=&4: G^{5/2}G^{3/2},  G^{7/2},  G^{5/2},
G^{3/2} \nonumber\\ \bar{m}(4)&\rightarrow&\bar{m}(4)-\bar{m}(0)-
\bar{m}(1/2)-
\bar{m}(3/2)-\bar{m}(5/2)           \nonumber\\ n&=&9/2:  G^{9/2},
G^{5/2}G^{3/2},  G^{7/2},  G^{5/2},
 G^{3/2} \nonumber\\
\bar{m}(9/2)&\rightarrow&\bar{m}(9/2)-\bar{m}(0)-\bar{m}(1/2)-
 \bar{m}(1)- \bar{m}(2)-\bar{m}(3)           \nonumber\\
\
\end{eqnarray}

\noindent The table of multiplity vectors for nonzero
norm states reads as:

\vl

\begin{tabular}{||c|c|c|c|c|c|c|c|c|c|c|c|c||}
 \hline\hline
n&1&8&10&$\bar{10}$&27&28&$\bar{28}$&35&$\bar{35}$&64&81&$\bar{81}$\\
\cline{1-13}\hline 0    &1&    &  &       &&&&&&&&\\ \hline
1/2&  &1  &  &       &&&&&&&&\\ \hline
1    &  &1  &1&1     &&&&&&&&\\ \hline
3/2&&2  &1&1   &1&&&&&&&\\ \hline
2    &1&3  &1&1   &3&&&&&&&\\ \hline
5/2&1&5  &2&2   &4&&&&&&&\\ \hline
3    &2&7&5&5&5 &&&2&2&1&&\\ \hline
7/2&2&12&6&6&9&&&         4   &4  &2&  &  \\ \hline
4    &6&16&9&9&16&    1&1&6  &6  &4&  &  \\ \hline
9/2&6&24&14&14&24&1&1&10&10&7&1&1\\ \hline \hline
\end{tabular}

\vl
\label{Nsnonzerodege}  Table 8.3. Table of multiplicity vectors for nonzero
norm N-S type color representations for various values of the conformal
weight $n$.

\vm

 This table contains all essential as regards to the construction of gauge
invariant states.  Assuming surjectivity for the maps induced by the
action of $G^{1/2}$,  $L^1$ and $L^2$ one has general expression for the
multiplicity vector of gauge invariant N-S type representations as

\begin{eqnarray}
\bar{m}(n)_{GI}&=& \bar{m}(n)-\bar{m}(n-1/2)-\bar{m}(n-1)-\bar{m}(n-2)
\end{eqnarray}

\noindent If some of the resulting multiplicities becomes negative it must
obviously be replaced with zero.  The  analogous formula for Ramond
representation is

\begin{eqnarray}
\bar{m}(n)_{GI}&=& \bar{m}(n)-2\bar{m}(n-1)-\bar{m}(n-1)-\bar{m}(n-2)
\end{eqnarray}

\noindent and obviously follows from the gauge conditions for $L^1$,$G^1$
and $L^2$.

\noindent In the following table the multiplicity vectors for gauge
 invariant states of N-S representations are listed and are used extensively
in calculations  of degeneracies for colored particles.

\vl

\begin{tabular}{||c|c|c|c|c|c|c|c|c|c|c|c|c||}
 \hline\hline
n&1&8&10&$\bar{10}$&27&28&$\bar{28}$&35&$\bar{35}$&64&81&$\bar{81}$\\
\cline{1-13}\hline 0    &1&    &  &       &&&&&&&&\\ \hline
1/2&  & 1 &  &       &&&&&&&&\\ \hline
1    &  &  &1&1     &&&&&&&&\\ \hline
3/2&  &  &  &   &1&&&&&&&\\ \hline
2  &  &  &  &   &2&&&&&&&\\ \hline
5/2&&  &   &   & &&&1&  1&&&\\ \hline
3    & &   &1& 1 &  & &     &1&1&1&&\\ \hline
7/2& &   &  &    &   &  &     &2&2&1&&\\ \hline
4    &1&  &   &  &   &  &      &1&1 &1 &  &  \\ \hline
9/2 & &   &   &  &  &   &     &   &  &1 &1&1\\ \hline \hline
\end{tabular}

\vl

\label{Nsginvdege}  Table 8.4.  Multiplicity vectors for N-S type color
representations satisfying Super Virasoro conditions as function of
conformal weight $n$. The table is needed in  the calculation of masses
of color excited  leptons and colored bosons.

\vm

The construction proceeds in similar manner for Ramond type algebra and the
 following table lists the results.

\vl

  \begin{tabular}{||c|c|c|c|c|c|c|c|c|c|c|c|c|c|c|c||}
 \hline\hline
n&1&8&10&$\bar{10}$&  27&28&$\bar{28}$&35&$\bar{35}$& 64&80&$\bar{80}$
&81&$\bar{81}$&125\\
\cline{1-16}\hline
 0    &1&    &  &       & &&&&&  &&&&&\\ \hline
1&  &     &  &       & &&&&&  &&&&&\\ \hline
2      &     & &1&1    &  &&&&&   &&&&&\\ \hline
3& & &&   &2&&&&& &&&&&\\ \hline
4 & &2 &  &   &    &  &  &1  &1  &  &&&&&\\ \hline
5&  & &2& 2 &  & &   & &  &1 &&&&&\\ \hline
6& &  &&  &  &1&1&&&2 &&&&&\\ \hline
7&& && &  &&&&&& & &1&1&\\ \hline
8&&&&&  1&&&4&4&1& & &1&1&1\\ \hline
9&&68&46&46&68&4&4&30&30&28&1&1&5&5&2\\ \hline \hline
\end{tabular}

\vl
\label{Raginvdege}  Table 8.5: Table of multiplicity vectors for nonzero
norm gauge invariant  Ramond type   color representations for various
values of the conformal
 weight $n$. Table is needed in calculation  mass of color excited states
 of quarks.

\section{ Appendix G:  Information on $so(4)$ and $u(1)$ type Super Virasoro
representations }

The following tables give the degeneracies  of operators creating nonzero
norm  states for $u(1)$ and $so..$ type representations.

\vl

\begin{tabular}{||c|c|c|c|c|c|c|c|c|c|c||}
 \hline\hline
(c,h) &$ \Delta$ &0&1/2&1&3/2&2&5/2&3&7/2&4\\ \cline{1-10}\hline
(3/2,1/2) &N &1&1&1&2&3&4&5&7&10\\ \hline
(3/2,0) & N&1&0&0&1&1&1&1&2&2\\ \hline\hline
\end{tabular}

\vl

\noindent \label{u1NSdege} Table 9.1. The number $N$ of operators for
$u(1)$ N-S type Super
 Virasoro representations  for central charge $c=3/2$ and  vacuum weight
$h= 1/2$ and $h=0$ as  function of the conformal weight $\Delta$.
In  $h=1/2$ case there are no singular vectors whereas in $h=0$ case the
operators $G^{1/2}$ and $L^1$ create zero norm states.

\vl

\begin{tabular}{||c|c|c|c|c|c|c|c|c|c|c||}
 \hline\hline
(c,h) &$ \Delta$ &0&1/2&1&3/2&2&5/2&3&7/2&4\\ \cline{1-10}\hline
(0,1/2) &N &1&1&1&2&3&4&5&7&8\\ \hline \hline
\end{tabular}

\vl

\noindent \label{NSdege} Table 9.2. The number $N$ of operators for
for $(c=0,h= 1/2)$ N-S representation as function of the conformal
weight $\Delta$.   For $(c=0,h=0)$ representation all states created
 by Super Virasoro generators
possess zero norm.

\vl

\begin{tabular}{||c|c|c|c|c|c|c||}
 \hline\hline
(c,h) &$ \Delta$ &0&1&2&3&4\\ \cline{1-7}\hline
(0,1/2) &N &2&2&4&8&14\\ \hline \hline
\end{tabular}

\vl

\noindent Table 9.3. \label{U1Radege} The number $N$ of operators for
for $(c=3/2, h=Q_K^2/2= 1/2)$ Ramond representation as function
 of the conformal
weight $\Delta$. The representation appears in fermionic
  $u(1)$ sector.
\vl

\begin{tabular}{||c|c|c|c|c|c|c||}
 \hline\hline
(c,h) &$ \Delta$ &0&1&2&3&4\\ \cline{1-7}\hline
(0,0) &N &1&2&4&8&14\\ \hline \hline
\end{tabular}

\vl

\noindent Table 9.4. \label{Rasodege} The number $N$ of operators    for
$(c=0,h= 0)$   $so(4)$ Ramond representation as function of the conformal
weight $\Delta$. The representation appears as basic building block of
 Kac Moody spinors.  Note that  $G^0$ creates
 zero norm state.

\section{ Appendix H: Number theoretic  auxiliary results}

The ground state degeneracies for fermions and bosons need not to be
identical
 to their ideal values $D=64$ and $D=16$ and it is of interest to find
under what conditions the degeneracy can be said to be near to its ideal
value.  This amounts to calculating the p-adic inverse of the $D$ in
general case. The calculation goes as follows.\\ a)  The problem  is to
find the lowest order term in p-adic expansion of the inverse $y$ of
p-adic number $x \in 1,...p-1$.  The remaining terms in expansion in
powers of $p$ can be found iteratively.   The equation to be solved is

\begin{eqnarray}
yx&=&1 \  mod \ p
\end{eqnarray}

\noindent for a given  value of $x $.  \\
b) One can express $p $ in the form

\begin{eqnarray}
p&=& Nx +r
\end{eqnarray}
\noindent The evaluation of $N$ and $r \in \{ 1,..,x-1 \}$ is a
 straightforward task.
The defining equation for $y$ can be written as

\begin{eqnarray}
yx&=& m(Nx +r) +1 =mNx+ mr+1
\end{eqnarray}

\noindent From this one must have

\begin{eqnarray}
mr+1 &=& kx
\end{eqnarray}

\noindent and any pair  $(m,k)$ satisfying this condition gives solution
 to  $y$:

\begin{eqnarray}
y&=& mN+ k
\end{eqnarray}

\noindent $y$ must be chosen to be the smallest possible one.

\vm

Consider as examples two practical  cases.  \\ a) $ p=M_{n}=2^{n}-1$ and
$x=15=2^4-1$.  One obtains   $r$ by   substituting repeatedly $2^4=1 \ mod
\ x $ to the expression
 of  $ M_{n}$. $M_{n}$ can be written in the form  $M_{n}= 15(2^{n-4}+
2^{n-8}+ ....) + r$ and the previous condition reads
 $mr+1= 15k$.\\ i)   For $M_{89}$ one has $r= 1$ and $(m,k)= (14,1) $
giving  $y=  14(2^{n-4}+2^{n-8} +...) +1$.  For the real counterpart  of
$Xp^2/2D$ one has the approximate expression $(7 X \ mod \  16)/15$ and
approximately N-S mass formula for small quantum numbers results.  \\ ii)
For $M_{127}$ and $M_{107}$ one has $r=7$ and $7m+1= 15k$ gives  $(m,k)=
2,1)$ and $y= 2(2^{n-4}+.....) + 1$. For $Xp^2/2D$ one has $ X mod 16/15$:
the factor $7$ is absent.     \\ b)$p=M^n$ and $x= 63=64-1$.  One
obtains   $r$  by   substituting repeatedly $2^6-1 \ mod \ x $ to the
expression of  $ M_{n}$.  One has  $r=1$ for $n=127$,  $r=31$ for $n=107$
and $n=89$.  For the real counterpart  $R$ of $Xp^2/D$ one has $ R=(62X\
mod \ 64)/(63M_n)$ and
 $y=(60 X \ mod \ 64)/(63M_n)$ for $n=127$ and $107,89$ respectively so
that mass formulas change somewhat and in $n$-dependent manner if one has
$D=63 $ instead of $D=64$. \\ b) $1/5$ factor appears in mass formulas for
leptons and the previous argument
 leads to the expression $p^2/5= (2^{126}-2^{124}+2^{122}-..)p^2$.  From
this formula the real counterpart of, say $2/5$,  is in good approximation
$4/5$.

\end{document}